\documentclass[10pt, journal]{IEEEtran}
\usepackage{graphicx,amssymb,amsmath}
\usepackage{multicol}
\usepackage[noadjust]{cite}
\usepackage{setspace}
\usepackage{subfigure}
\usepackage{graphicx}
\usepackage{float}
\usepackage{url}
\usepackage{stfloats}
\usepackage{amsthm,pifont}
\usepackage{flushend}
\usepackage{cases,subeqnarray}
\usepackage{bm,multirow,bigstrut}
\usepackage{amsmath, amsthm, amssymb}
\usepackage{textcomp}
\usepackage{latexsym,bm}
\usepackage{booktabs}
\usepackage{xcolor}
\usepackage{mathtools}
\usepackage{dsfont}
\usepackage{extarrows}
\usepackage{epsfig}
\usepackage{epsfig}
\usepackage{epstopdf}
\usepackage[noend]{algpseudocode}
\usepackage{algorithmicx,algorithm}
    \usepackage{makecell}
    	
		\theoremstyle{plain}

		\theoremstyle{plain}
		\newtheorem{rem}{Remark}
		\newtheorem{coro}{Corollary}
		\newtheorem{prop}{Proposition}
		\usepackage{amsmath}
		\newtheorem{theorem}{Theorem}
		
		\IEEEoverridecommandlockouts
		
\begin{document}
		\title{Secrecy Performance of Body-Centric Communications over Alternate Rician Shadowed Fading Channels}
		\author{Xiaoqi Wang, Jiayi~Zhang,~\IEEEmembership{Senior~Member,~IEEE}, Hongyang Du, Ning Wang, and \\Bo Ai,~\IEEEmembership{Senior~Member,~IEEE}
		\thanks{X. Wang, J.~Zhang, and H. Du are with the School of Electronic and Information Engineering, Beijing Jiaotong University, Beijing 100044, P. R. China. (e-mail: jiayizhang@bjtu.edu.cn)}
		\thanks{N. Wang is with School of Information Engineering, Zhengzhou University, Zhengzhou 450001, China.}
		\thanks{B. Ai is with State Key Laboratory of Rail Traffic Control and Safety, Beijing Jiaotong University, Beijing 100044, China.}
		}
				
		\maketitle
		\begin{abstract}			
In this paper, we investigate the physical layer security over the Alternate Rician Shadowed fading channel, which is a novel channel model for body-centric wireless links and land mobile satellite. We derive exact closed-form expressions for the average secrecy capacity (ASC), secrecy outage probability (SOP), and probability of non-zero secrecy capacity (PNZ) for two cases: (i) $m$ is a positive real number and (ii) $m$ is a positive integer number, where $m$ describes the level of fluctuation of the line-of-sight component. In the first case, SOP is derived in terms of the Meijer's $G$-function, while ASC and PNZ are derived in terms of the multivariate Fox's $H$-function. In the second case, ASC is derived in terms of the Meijer's $G$-function, while SOP and PNZ are derived in terms of elementary functions.
In addition, we derive the asymptotic ASC, SOP and PNZ expressions which all match well the exact ones at high values of  signal-to-noise ratio, respectively.
The capacity slope of asymptotic ASC and the secrecy diversity
order of asymptotic SOP have been
derived for providing more physical insights.
Finally, the accuracy of our derived expressions is validated by Monte-Carlo simulations.
		\end{abstract}
		\begin{IEEEkeywords}
 Alternate Rician Shadowed fading channels, Fox's $ H $-function, physical layer security.
		\end{IEEEkeywords}
		\IEEEpeerreviewmaketitle
		\section{Introduction}
		Physical layer security (PLS) has been initially proposed in
		\cite{wyner1975wire} as a promising solution to strengthen the secure transmission of wireless communications
		by using the information-theoretic approach.
		Body-centric communications \cite{cotton2013statistical} is one of the wireless applications of emerging interests, where at least one of the system transceivers is physically located on the body of a person, which acts as a final node or as a relay in a network by employing device-to-device (D2D) communications \cite{cotton2014human}. Due to the highly standardized communication scheme, it is increasingly vulnerable for legitimate D2D pairs to highly ensure secrecy from malicious third entities, especially when they are being wiretapped due to the open access of transmission media \cite{ shiu2011physical}.
		
		In order to model the  body-centric fading channels, the Alternate Rician Shadowed (ARS) distribution has been proposed in \cite{fernandez2019tractable} as a tractable and important fading model which consists of two fluctuating specular components of which only one is active at a time. It can be regarded as a mixture of two Rician shadowed fading models, when a diffuse component is added. This distribution can be reduced to other common fading models, such as the classical Rayleigh, Rician, Nakagami-$ m $ and Hoyt fading channels. Furthermore, the authors in  \cite{fernandez2019tractable} proved that the ARS model can provide either left-sided or right-sided bimodality (i.e. two-sided bimodality), depending on its shape parameters, which is not found in classical fading models, or even in more recent ones such as the Two Wave with Diffuse Power  and  Fluctuating Two Ray (FTR) models. In addition, the ARS model provides a better fit to the experimental measurements performed in the body-centric wireless system \cite{cotton2013statistical }, especially in comparison with the FTR model \cite{romero2017fluctuating,8727983,8085132} and  the $\kappa$-$\mu$ /inverse Gamma composite fading model \cite{Yoo2017The}. Due to its promising properties, ARS model is suitable for fitting the bimodal fading distributions measured in some typical communications systems such as body-centric wireless links and land mobile satellite.
	
	On the other hand, the performance of the secure communications over generalized fading channels has been widely investigated in the open literature.
	In \cite{ zeng2018physical }, the average secrecy capacity (ASC), the secrecy outage probability (SOP), and the probability of non-zero secrecy capacity (PNZ)  over the FTR fading channels were derived. In addition, the effective capacity of a wireless communication system operating in the presence of FTR fading channels were given in \cite{ peppas2018effective }.
	In \cite{an2016secrecy}, SOP and PNZ over the Shadowed-Rician fading channels were derived. The probability of positive secrecy capacity and the upper bound of SOP over $ \alpha  $-$ \mu  $ fading channels were derived in \cite{kong2015performance}.
	The ASC and the SOP over $\kappa {\rm{ - }}\mu $ shadowed fading are given in \cite{ lopez2016calculation } for integer fading parameters. This channel model is also utilised in \cite{sun2018performance } to analyse the lower bound of SOP and the probability of strictly positive secrecy capacity using the exact probability density function (PDF).
	 However, the PLS of wireless communications
	over ARS fading channels has not been investigated in previous works.
		
		Motivated by the experimental and theoretical advantages of ARS distribution in body-centric wireless links and land mobile satellite systems, we provide an investigation on the PLS over the ARS fading channels\footnote{The presented method can be extended to multiple-antenna systems.}. The main contributions of this paper are summarized as follows:
		\begin{itemize}
		\item We analyze the performance of the PLS when both main and wiretap channels are subjected to ARS fading channel models. Considering the presence of active and passive eavesdroppers, we derive three essential secrecy metrics with exact closed-form expressions, including SOP, PNZ, and ASC. More specifically SOP is derived in terms of Meijer's $ G $-functions, while ASC and PNZ are exactly derived in terms of multivariate Fox's $H$-functions. Note that the multivariate Fox's $H$-functions can be efficiently implemented by various programming codes, such as the Python code in \cite{ alhennawi2015closed} and  the Matlab code in \cite{chergui2019rician}.
		\item We derive the asymptotic expressions of ASC, SOP and PNZ in the high-SNR regime. The asymptotic results demonstrate that the secrecy diversity
		order (SDO) of asymptotic SOP and the capacity slope of asymptotic ASC equal constants.
		 These asymptotic expressions all well match the exact ones at high values of signal-to-noise ratio (SNR).
		\item To provide more physical insights into the performance of the PLS as well as the impact of the parameters of ARS fading model, we derive simple and useful performance metrics, including SOP, PNZ and ASC, when the fading parameter $ m $  is assumed to be integer values.
		\end{itemize}
		
The remainder of the paper is organized as follows. In Section II, we introduce the statistical characterizations of the ARS distribution, and present PDF and cumulative distribution function (CDF) of ARS fading channels for two cases: (i) $ m $ is positive real number and (ii) $ m $ is a positive integer number. The closed-form ASC, SOP and PNZ expressions are derived in Section III, Section IV and Section V, respectively. The Monte-Carlo simulations, numerical results and discussions are subsequently presented in Section VI. Finally, Section VII concludes this paper.

		\section{System And Channel Models}
		
		\subsection{System Model}
\begin{figure}[t]
			\centering
			\includegraphics[scale=0.31]{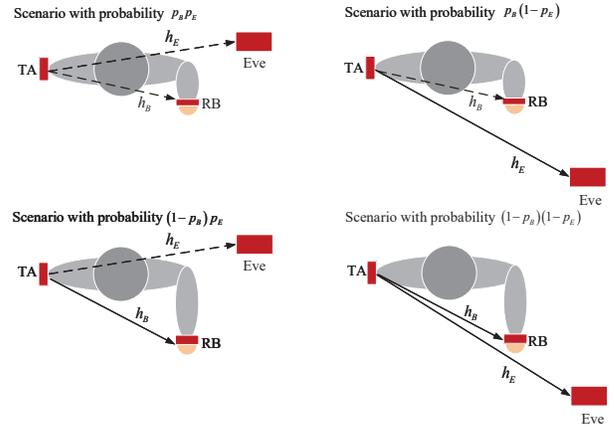}
			\caption{Illustration of system model with two legitimate transceivers	(TA and RB) and one eavesdropper (Eve).}
			\label{SYSTEM MODEL}
		\end{figure}

	Fig. \ref{SYSTEM MODEL} shows a possible ARS fading scenario where the transmitter (TA) is located at the right-waist of the person and the legitimate receiver (RB) is located at the left-wrist corresponding to a body-centric system configuration. The natural movement of the arms leads to two prominent states of shadowing of the main LoS signal, which result in the bimodal characteristic of the statistical model.
	
	There are three nodes with two wireless communication links in the Wyner’s wiretap channel \cite{wyner1975wire} as shown in Fig. \ref{SYSTEM MODEL}. The link between TA and RB is the main channel, while the other wireless communication link describes the wiretap channel between TA and an eavesdropper (Eve). RB's channel state information (CSI) is assumed to be known by TA, while Eve’s accurate CSI is unknown. Thus, the information-theoretic security cannot be guaranteed.
	
		In this paper, both the main and wiretap channels, $ {h_\ell }$, $\ell  \in \left\{ {B,E} \right\} $ are assumed to be modeled by the ARS fading model \cite{fernandez2019tractable}. In addition, we assume that TA, RB and Eve are equipped with a single antenna.
		
		The received instantaneous SNR at RB and Eve can be  expressed as
		{\small\begin{equation}\label{1}
		{\gamma _\ell } = \frac{{{X_\ell }}}{{{\sigma ^2}}} = \frac{{\left| {V_\ell ^2} \right|}}{{{\sigma ^2}}},
		\end{equation}}
		where ${X_\ell }$ ($\ell  \in \left\{ {B,E} \right\}$) is the received signal power envelope. $B$ and $E$ respectively stand for the legitimate receiver RB and Eve, ${V_\ell }$ is the complex baseband signal at the receiver, and  $\sigma ^2$ is the noise variance.
		The complex baseband signal ${V_\ell }$ can be expressed as
		{\small\begin{equation}\label{2}
		{V_\ell } = A{V_{\ell ,1}} + \left( {1 - A} \right){V_{\ell ,2}} + {G_1} + j{G_2},
		\end{equation}}
		where $j=\sqrt{-1}$, $A$ is a Bernoulli random variable which takes the value one with the probability $p$ and the value zero with the probability $1-p$,  ${V_{\ell ,r}} (r \in \left\{ {1,2} \right\}$)  represents the first and second possible line-of-sight (LoS) components, respectively, and ${G_1} + j{G_2}$ is a complex circularly symmetric Gaussian random variable which represents the diffuse received signal component.
		
		According to \eqref{2}, the ARS models a channel where two alternative Rician shadowed types of fading can be observed but only one at a time, and each one has its Rician parameter and total power as
		{\small\begin{align}\label{3}
		{K_{\ell ,r}} = \frac{{\left| {V_{\ell ,r}^2} \right|}}{{2{\sigma ^2}}},	
		\end{align}
		\begin{align}\label{4}
		{\Omega _{\ell ,r}} = \left| {V_{\ell ,r}^2} \right| + 2{\sigma ^2}.\!\!\!\!\!\!\!\!\!\!\!\!\!\!\!
		\end{align}}
		In order to facilitate the analysis later, the average power and average Rician factor are defined as
		{\small\begin{equation}\label{5}
		{\overline K _\ell }\buildrel \Delta \over = {p_\ell }{K_{\ell ,1}} + \left( {1 - {p_\ell }} \right){K_{\ell ,2}},
		\end{equation}
		\begin{equation}\label{6}
		{\overline \Omega  _\ell } \buildrel \Delta \over = {p_\ell }{\Omega _{\ell ,1}} + \left( {1 - {p_\ell }} \right){\Omega _{\ell ,2}}.
		\end{equation}}
		\subsection{The PDF and CDF of Alternate Rician Shadowed Fading Model}
		The PDF and CDF of ARS power envelope $X$ have been given by \cite[Eq. (10)]{fernandez2019tractable}. By applying the simple change of variables $\gamma  = {\raise0.7ex\hbox{$X$} \!\mathord{\left/
				{\vphantom {X {{\sigma ^2}}}}\right.\kern-\nulldelimiterspace}
			\!\lower0.7ex\hbox{${{\sigma ^2}}$}} = {\raise0.7ex\hbox{${\overline \gamma  X}$} \!\mathord{\left/
				{\vphantom {{\overline \gamma  X} {\overline \Omega  }}}\right.\kern-\nulldelimiterspace}
			\!\lower0.7ex\hbox{${\overline \Omega  }$}}$ with the single power envelop X, the PDF and CDF of the instantaneous SNR  can be expressed as
		{\small  \begin{align}\label{7}
		{f_\ell }\left( \gamma  \right)=&{p_\ell }{f_{RS,\ell ,1}}\left( {\gamma ;\frac{{{{\overline \gamma  }_\ell }\left( {1 + {K_{\ell ,1}}} \right)}}{{1 + {{\overline K }_\ell }}},{K_{\ell ,1}},{m_\ell }} \right)\notag\\
		& + \left( {1 - {p_\ell }} \right){f_{RS,\ell ,2}}\left( {\gamma ;\frac{{{{\overline \gamma  }_\ell }\left( {1 + {K_{\ell ,2}}} \right)}}{{1 + {{\overline K }_\ell }}},{K_{\ell ,2}},{m_\ell }} \right),
		\end{align}}
		{\small\begin{align}\label{8}
		{F_\ell }\left( \gamma  \right)=&{p_\ell }{F_{RS,\ell ,1}}\left( {\gamma ;\frac{{{{\overline \gamma  }_\ell }\left( {1 + {K_{\ell ,1}}} \right)}}{{1 + {{\overline K }_\ell }}},{K_{\ell ,1}},{m_\ell }} \right) \notag \\
		&+ \left( {1 - {p_\ell }} \right){F_{RS,\ell ,2}}\left( {\gamma ;\frac{{{{\overline \gamma  }_\ell }\left( {1 + {K_{\ell ,2}}} \right)}}{{1 + {{\overline K }_\ell }}},{K_{\ell ,2}},{m_\ell }} \right),
		\end{align}}
		where ${\overline \gamma  _\ell }$ is the average SNR, ${m_\ell }$  is the shadowing parameter, ${p_\ell }$  is the possibility of observing the Rician shadowed distribution with parameters ${\Omega _{\ell ,1}}$, ${K_{\ell ,1}}$, and ${m_\ell }$, $1-{p_\ell }$ is the probability of observing the ARS distribution with parameters  ${\Omega _{\ell ,2}}$, ${K_{\ell ,2}}$, and ${m_\ell }$, ${f_{RS,\ell ,r}}\left( {\gamma ;\frac{{{{\overline \gamma  }_\ell }\left( {1 + {K_{\ell ,r}}} \right)}}{{1 + {{\overline K }_\ell }}},{K_{\ell ,r}},{m_\ell }} \right)$  and ${F_{RS,\ell ,r}}\left( {\gamma ;\frac{{{{\overline \gamma  }_\ell }\left( {1 + {K_{\ell ,r}}} \right)}}{{1 + {{\overline K }_\ell }}},{K_{\ell ,r}},{m_\ell }} \right)$  are the PDF and CDF of the Rician shadowed instantaneous SNR $\gamma$. In this paper, they are the same with ${f_{RS,\ell ,r}}\left( \gamma  \right)$  and ${F_{RS,\ell ,r}}\left( \gamma  \right)$, respectively, for convenience.

		 When $m$ is a  positive real number, the PDF and CDF of the Rician shadowed power envelope $X$ are given as \cite[Eq. (8)]{fernandez2019tractable}. By applying the same change of variables $\gamma  = {\raise0.7ex\hbox{$X$} \!\mathord{\left/
				{\vphantom {X {{\sigma ^2}}}}\right.\kern-\nulldelimiterspace}
			\!\lower0.7ex\hbox{${{\sigma ^2}}$}} = {\raise0.7ex\hbox{${\overline \gamma  X}$} \!\mathord{\left/
				{\vphantom {{\overline \gamma  X} {\overline \Omega  }}}\right.\kern-\nulldelimiterspace}
			\!\lower0.7ex\hbox{${\overline \Omega  }$}}$, we can obtain 	
	 {\small \begin{align}\label{9}
		&{f_{RS,\ell ,r}}\!\left( \!{\gamma ;\frac{{{{\overline \gamma  }_\ell }\left( {1 + {K_{\ell ,r}}} \right)}}{{1 + {{\overline K }_\ell }}},{K_{\ell ,r}},{m_\ell }} \!\right)
		\!=\! \frac{{1 + {{\overline K }_\ell }}}{{{{\overline \gamma  }_\ell }}}{\left( {\frac{{{m_\ell }}}{{{m_\ell } + {K_{\ell ,r}}}}} \right)^{{m_\ell }}}\!\!\!\!\!\notag\\
		&\times \exp \left( { - \frac{{\left( {1 + {{\overline K }_\ell }} \right)}}{{{{\overline \gamma  }_\ell }}}\gamma } \right)\!{}_1{F_1}\left( {{m_\ell },1;\frac{{{K_{\ell ,r}}\left( {1 + {{\overline K }_\ell }} \right)}}{{{{\overline \gamma  }_\ell }\left( {{m_\ell } + {K_{\ell ,r}}} \right)}}\gamma } \right),
		\end{align}}
		 {\small \begin{align}\label{10}
		&{F_{RS,\ell ,r}}\!\left(\! {\gamma ;\frac{{{{\overline \gamma  }_\ell }\left( {1 + {K_{\ell ,r}}} \right)}}{{1 + {{\overline K }_\ell }}},{K_{\ell ,r}},{m_\ell }}\! \right) \!\!=\!\! \frac{{1 + {{\overline K }_\ell }}}{{{{\overline \gamma  }_\ell }}}{\left( {\frac{{{m_\ell }}}{{{m_\ell } + {K_{\ell ,r}}}}} \right)^{{m_\ell }}}\!\!\notag\\
		&\times\gamma {\Phi _2}\left( {1 - {m_\ell },{m_\ell };2; - \frac{{1 + {{\overline K }_\ell }}}{{{{\overline \gamma  }_\ell }}}\gamma , - \frac{{{m_\ell }\left( {1 + {{\overline K }_\ell }} \right)}}{{{{\overline \gamma  }_\ell }\left( {{m_\ell } + {K_{\ell ,r}}} \right)}}\gamma } \right).
		\end{align}}
		By substituting (\ref{9}) and (\ref{10}) into (\ref{7}) and (\ref{8}), respectively, the PDF and CDF of the SNR over the ASR model can be expressed as
		{\small \begin{align}\label{11}
		&{f_\ell }\left( \gamma  \right)\! =\! {p_\ell }\frac{{1 \!+\! {{\overline K }_\ell }}}{{{{\overline \gamma  }_\ell }}}{\left(\! {\frac{{{m_\ell }}}{{{m_\ell }\! +\! {K_{\ell ,1}}}}} \!\right)^{{m_\ell }}}\!\!\!{}_1{F_1}\left( {{m_\ell },1;\frac{{{K_{\ell ,1}}\left( {1\! + \!{{\overline K }_\ell }} \right)}}{{{{\overline \gamma  }_\ell }\left( {{m_\ell }\! +\! {K_{\ell ,1}}} \right)}}\gamma } \right)\notag\\
		&\times
		\exp \left( { - \frac{{\left( {1 + {{\overline K }_\ell }} \right)}}{{{{\overline \gamma  }_\ell }}}\gamma } \right)
		+ \left( {1 - {p_\ell }} \right)\frac{{1 + {{\overline K }_\ell }}}{{{{\overline \gamma  }_\ell }}}{\left( {\frac{{{m_\ell }}}{{{m_\ell } + {K_{\ell ,2}}}}} \right)^{{m_\ell }}} \notag\\
		&\times{}_1{F_1}\left( {{m_\ell },1;\frac{{{K_{\ell ,2}}\left( {1 + {{\overline K }_\ell }} \right)}}{{{{\overline \gamma  }_\ell }\left( {{m_\ell } + {K_{\ell ,2}}} \right)}}\gamma } \right)\exp \left( { - \frac{{\left( {1 + {{\overline K }_\ell }} \right)}}{{{{\overline \gamma  }_\ell }}}\gamma } \right),
		\end{align}	}
		{\small \begin{align}\label{12}
		{F_\ell }\left( \gamma  \right)\!& = {p_\ell }\frac{{1 + {{\overline K }_\ell }}}{{{{\overline \gamma  }_\ell }}}{\left( {\frac{{{m_\ell }}}{{{m_\ell } + {K_{\ell ,1}}}}} \right)^{{m_\ell }}} \gamma\notag\\
		&\times\!{\Phi _2}\!\left(\! {1 - {m_\ell },{m_\ell };2; - \frac{{1 + {{\overline K }_\ell }}}{{{{\overline \gamma  }_\ell }}}\gamma\! , -\! \frac{{{m_\ell }\left( {1 + {{\overline K }_\ell }} \right)}}{{{{\overline \gamma  }_\ell }\left( {{m_\ell } + {K_{\ell ,1}}} \right)}}\gamma } \right) \notag\\
		& + \left( {1 - {p_\ell }} \right)\frac{{1 + {{\overline K }_\ell }}}{{{{\overline \gamma  }_\ell }}}{\left( {\frac{{{m_\ell }}}{{{m_\ell } + {K_{\ell ,2}}}}} \right)^{{m_\ell }}}\!\gamma \notag\\
		&\times\!{\Phi _2}\!\left( \!{1\! -\! {m_\ell },{m_\ell };2; - \frac{{1 + {{\overline K }_\ell }}}{{{{\overline \gamma  }_\ell }}}\gamma\! , -\! \frac{{{m_\ell }\left( {1 \!+\! {{\overline K }_\ell }} \right)}}{{{{\overline \gamma  }_\ell }\left( {{m_\ell } + {K_{\ell ,2}}} \right)}}\gamma } \right)\!\!,
		\end{align}	}
		where ${}_1{F_1}\left( \cdot \right)$  is the confluent hypergeometric function \cite[Eq. (9.210.1)]{gradshteyn2007}, and ${\Phi _2}\left( \cdot \right)$  is the bivariate confluent hypergeometric function \cite[Eq. (1.4.8)]{srivastava1985multiple}.
		
		 When $m$ is a positive integer number, the PDF and CDF of the Rician shadowed power envelope $X$ are given as \cite[Eq.  (9)]{fernandez2019tractable}.  Employing similar transformation,
		the PDF and CDF of the SNR over the ASR model can be expressed as
{\small 		\begin{align}\label{15}
		&{f_\ell }\left( \gamma  \right) = {p_\ell }\frac{{1 + {{\overline K }_\ell }}}{{{{\overline \gamma  }_\ell }}}{\left( {\frac{{{m_\ell }}}{{{m_\ell } + {K_{\ell ,1}}}}} \right)^{{m_\ell }}}\exp \left( {\frac{{ - \left( {1 + {{\overline K }_\ell }} \right){m_\ell }}}{{\left( {{m_\ell } + {K_{\ell ,1}}} \right){{\overline \gamma  }_\ell }}}\gamma } \right)\notag\\
		&\quad\times{L_{{m_\ell } - 1}}\left( { - \frac{{{K_{\ell ,1}}\left( {1 + {{\overline K }_\ell }} \right)}}{{{{\overline \gamma  }_\ell }\left( {{m_\ell } + {K_{\ell ,1}}} \right)\left( {1 + {K_{\ell ,1}}} \right)}}\gamma } \right)\notag\\
		& + \left( {1 - {p_\ell }} \right)\frac{{1 + {{\overline K }_\ell }}}{{{{\overline \gamma  }_\ell }}}{\left( {\frac{{{m_\ell }}}{{{m_\ell } + {K_{\ell ,2}}}}} \right)^{{m_\ell }}}\exp \left( {\frac{{ - \left( {1 + {{\overline K }_\ell }} \right){m_\ell }}}{{\left( {{m_\ell } + {K_{\ell ,2}}} \right){{\overline \gamma  }_\ell }}}\gamma } \right)\notag\\
		&\quad\times{L_{{m_\ell } - 1}}\left( { - \frac{{{K_{\ell ,1}}\left( {1 + {{\overline K }_\ell }} \right)}}{{{{\overline \gamma  }_\ell }\left( {{m_\ell } + {K_{\ell ,2}}} \right)\left( {1 + {K_{\ell ,2}}} \right)}}\gamma }\! \right),
		\end{align}}	
		{\small \begin{align}\label{16}
		&{F_\ell }\left( \gamma  \right) = 1 - {p_\ell }\frac{{{m_\ell } + {K_{\ell ,1}}}}{{{m_\ell }}}\exp \left( {\frac{{ - \left( {1 + {{\overline K }_\ell }} \right){m_\ell }}}{{\left( {{m_\ell } + {K_{\ell ,1}}} \right){{\overline \gamma  }_\ell }}}\gamma } \right)\notag\\
		&\quad\times\sum\limits_{k = 0}^{{m_\ell } - 1} {{{\left( {\frac{{{K_{\ell ,1}}}}{{{m_\ell }}}} \right)}^k}L_{{m_\ell } - k - 1}^k\left( {\frac{{ - \left( {1 + {{\overline K }_\ell }} \right){K_{\ell ,2}}}}{{\left( {{m_\ell } + {K_{\ell ,1}}} \right){{\overline \gamma  }_\ell }}}\gamma } \right)} \notag\\
		& - \left( {1 - {p_\ell }} \right)\frac{{{m_\ell } + {K_{\ell ,2}}}}{{{m_\ell }}}\exp \left( {\frac{{ - \left( {1 + {{\overline K }_\ell }} \right){m_\ell }}}{{\left( {{m_\ell } + {K_{\ell ,2}}} \right){{\overline \gamma  }_\ell }}}\gamma } \right)\notag\\
		&\quad\times\sum\limits_{k = 0}^{{m_\ell } - 1} {{{\left( {\frac{{{K_{\ell ,2}}}}{{{m_\ell }}}} \right)}^k}L_{{m_\ell } - k - 1}^k\left( {\frac{{ - \left( {1 + {{\overline K }_\ell }} \right){K_{\ell ,2}}}}{{\left( {{m_\ell } + {K_{\ell ,2}}} \right){{\overline \gamma  }_\ell }}}\gamma } \right)},
		\end{align}}
		where $L_n^m\left( \cdot \right)$  is the Generalized Laguerre Polynomials [20].
		\section{Average Secrecy Capacity}
		
		In this section, we derive exact expressions, integer approximate expressions, and asymptotic expressions in the high-SNR regime for ASC. Based on the asymptotic expressions, we characterize ASC in terms of the high-SNR slope to explicitly
		capture the impact of channel parameters on ASC performance at high SNRs \cite{wang2014secure}.
		\subsection{Exact and Integer Approximate ASC}
		The ASC can be expressed as \cite[Eq. (17)]{lei2015physical}
		{\small\begin{equation}\label{17}
		{\overline C _{\rm s}} = {I_1} + {I_2} - {I_3},
		\end{equation}}
		where
		{\small\begin{equation}\label{18}
		{I_1}  \buildrel \Delta \over =  \int_0^\infty  {\ln \left( {1 + {\gamma _B}} \right){f_B}\left( {{\gamma _B}} \right){F_E}\left( {{\gamma _B}} \right)d{\gamma _B}},
		\end{equation}
		\begin{equation}\label{19}
		{I_2} \buildrel \Delta \over = \int_0^\infty  {\ln \left( {1 + {\gamma _E}} \right){f_E}\left( {{\gamma _E}} \right){F_B}\left( {{\gamma _E}} \right)d{\gamma _E}},
		\end{equation}
		\begin{equation}\label{20}
		\!\!\!\!\!\!\!\!\!\!\!\!\!\!\!\!\!\!\!{I_3} \buildrel \Delta \over = \int_0^\infty  {\ln \left( {1 + {\gamma _E}} \right){f_E}\left( {{\gamma _E}} \right)d{\gamma _E}}.
		\end{equation}}
				
		\begin{theorem}\label{prop1}
		For a real value of $ m $, exact closed-form expressions for ${I_1}$, ${I_2}$ and ${I_3}$
		can be expressed as
		{\small\begin{equation}\label{21}
		{I_1} = {R_{1,1}} + {R_{1,2}} + {R_{2,1}} + {R_{2,2}},
		\end{equation}
		\begin{equation}\label{22}
		{I_2} = {T_{1,1}} + {T_{1,2}} + {T_{2,1}} + {T_{2,2}},
		\end{equation}}
		and
		\eqref{23} at the bottom of the next page, 	
		where $ H_{p,q:{p_1},{q_1},...,{p_L},{q_L}}^{0,n:{m_1},{n_1},...,{m_L},{n_L}}\left[ \cdot \right] $ is the multivariate Fox's $H$-function \cite[Eq. (28)]{alhennawi2015closed},
		$ {R_{i,j}} $ and $ {T_{i,j}} $ are derived as
		\eqref{24} and \eqref{25} at the bottom of the next page,
		and		
		\newcounter{mytemp1}
		\begin{figure*}[b]
			\normalsize
			\setcounter{mytemp1}{\value{equation}}
			\hrulefill
			\setcounter{equation}{20}	
				{\small \begin{align}\label{23}
					{I_3}& = \frac{{{p_E}}}{{\Gamma \left( {1 - {m_E}} \right)}}{\left( {\frac{{{m_E}}}{{{m_E} + {K_{1,E}}}}} \right)^{{m_E}{\rm{ - 1}}}} \!\!\!\!\!\!H_{1,2:1,1;1,2}^{0,1:1,1;2,1}\left[ {\frac{{{K_{1,E}}}}{{{m_E}}},\frac{{{m_E}\left( {1 + {{\overline K }_E}} \right)}}{{{{\overline \gamma  }_E}\left( {{m_E} + {K_{1,E}}} \right)}}\left|\!\!\! {\begin{array}{*{20}{c}}
							{\left( {0;1, - 1} \right)}\\
							{\left( {0;1,0} \right),\left( {0;0, - 1} \right)}
							\end{array}\!\!\!\begin{array}{*{20}{c}}
							:\\
							:
							\end{array}\!\!\!\begin{array}{*{20}{c}}
							{\left( {{m_E},1} \right)}\\
							{\left( {0,1} \right)}
							\end{array}\!\!\!\begin{array}{*{20}{c}}
							;\\
							;
							\end{array}\!\!\!\begin{array}{*{20}{c}}
							{\left( {0,1} \right)}\\
							{\left( {0,1} \right),\left( {0,1} \right)}
							\end{array}} \right.} \!\!\!\right]\notag\\
					&{\rm{ + }}\frac{{1 \!-\! {p_E}}}{{\Gamma \left( {1 - {m_E}} \right)}}{\left( {\frac{{{m_E}}}{{{m_E}\! +\! {K_{{\rm{2}},E}}}}} \right)^{{m_E}{\rm{ - 1}}}}\!\!\! H_{1,2:1,1;2,1}^{0,1:1,1;1,2}\!\!\left[ {\frac{{{K_{{\rm{2}},E}}}}{{{m_E}}},\frac{{{m_E}\!\left( {1 \!+\! {{\overline K }_E}} \right)}}{{{{\overline \gamma  }_E}\!\left( {{m_E}\! +\! {K_{1,E}}} \right)}}\!\left|\! {\begin{array}{*{20}{c}}
							{\left( {0;1, - 1} \right)}\\
							{\left( {0;1,0} \right),\left( {0;0, - 1} \right)}
							\end{array}\!\!\!\begin{array}{*{20}{c}}
							:\\
							:
							\end{array}\!\!\!\begin{array}{*{20}{c}}
							{\left( {{m_E},1} \right)}\\
							{\left( {0,1} \right)}
							\end{array}\!\!\!\begin{array}{*{20}{c}}
							;\\
							;
							\end{array}\!\!\!\begin{array}{*{20}{c}}
							{\left( {0,1} \right)}\\
							{\left( {0,1} \right),\left( {0,1} \right)}
							\end{array}} \right.} \right],
					\end{align} }
				\hrulefill
				{\small \begin{align}\label{24}
					{R_{i,j}} \buildrel \Delta \over =&{{Q}_{B,i}}{Q_{E,j}}\frac{{{{\overline \gamma  }_B}\left( {1 + {{\overline K }_E}} \right)}}{{{{\overline \gamma  }_E}\left( {1 + {{\overline K }_B}} \right)}}{\left( {\frac{{{m_E}}}{{{m_E} + {K_{E,j}}}}} \right)^{{m_E}}}{\left( {\frac{{{m_B}}}{{{m_B} + {K_{B,i}}}}} \right)^{{m_B} - 2}}\frac{1}{{\Gamma \left( {1 - {m_B}} \right)\Gamma \left( {1 - {m_E}} \right)\Gamma \left( {{m_E}} \right)}}\notag\\
					&\times H_{1,3:1,1;1,1;1,1;2,1}^{0,1:1,1;1,1;1,1;1,2}\left[ {\left. {\frac{{{{\overline \gamma  }_B}\left( {{m_B} + {K_{B,i}}} \right)\left( {1 + {{\overline K }_E}} \right)}}{{{{\overline \gamma  }_E}{m_B}\left( {1 + {{\overline K }_B}} \right)}},\frac{{{{\overline \gamma  }_B}{m_E}\left( {{m_B} + {K_{B,i}}} \right)\left( {1 + {{\overline K }_E}} \right)}}{{{{\overline \gamma  }_E}{m_B}\left( {{m_E} + {K_{E,j}}} \right)\left( {1 + {{\overline K }_B}} \right)}},\frac{{{K_{B,i}}}}{{{m_B}}},\frac{{{{\overline \gamma  }_B}\left( {{m_B} + {K_{B,i}}} \right)}}{{{m_B}\left( {1 + {{\overline K }_B}} \right)}}} \right|} \right.\notag\\
					&\left. {\begin{array}{*{20}{c}}
						{\left( { - 1;1,1,1,1} \right)}\\
						{\left( { - 1;1,1,0,0} \right),\left( {0;0,0,1,0} \right),\left( {0;0,0,0,1} \right)}
						\end{array}\begin{array}{*{20}{c}}
						:\\
						:
						\end{array}\begin{array}{*{20}{c}}
						{\left( {{m_E},1} \right)}\\
						{\left( {0,1} \right)}
						\end{array}\begin{array}{*{20}{c}}
						;\\
						;
						\end{array}\begin{array}{*{20}{c}}
						{\left( {1 - {m_E},1} \right)}\\
						{\left( {0,1} \right)}
						\end{array}\begin{array}{*{20}{c}}
						;\\
						;
						\end{array}\begin{array}{*{20}{c}}
						{\left( {{m_B},1} \right)}\\
						{\left( {0,1} \right)}
						\end{array}\begin{array}{*{20}{c}}
						;\\
						;
						\end{array}\begin{array}{*{20}{c}}
						{\left( {1,1} \right),\left( {1,1} \right)}\\
						{\left( {1,1} \right)}
						\end{array}} \right],
					\end{align} }			
				\hrulefill
	{\small \begin{align}\label{25}
		{T_{i,j}} \buildrel \Delta \over =&{{\rm{Q}}_{E,i}}{Q_{B,j}}\frac{{{{\overline \gamma  }_E}\left( {1 + {{\overline K }_B}} \right)}}{{{{\overline \gamma  }_B}\left( {1 + {{\overline K }_E}} \right)}}{\left( {\frac{{{m_B}}}{{{m_B} + {K_{B,j}}}}} \right)^{{m_B}}}{\left( {\frac{{{m_E}}}{{{m_E} + {K_{E,i}}}}} \right)^{{m_E} - 2}}\frac{1}{{\Gamma \left( {1 - {m_E}} \right)\Gamma \left( {1 - {m_B}} \right)\Gamma \left( {{m_B}} \right)}}\notag\\
		&\times H_{1,3:1,1;1,1;1,1;2,1}^{0,1:1,1;1,1;1,1;1,2}\left[ {\left. {\frac{{{{\overline \gamma  }_E}\left( {{m_E} + {K_{E,i}}} \right)\left( {1 + {{\overline K }_B}} \right)}}{{{{\overline \gamma  }_B}{m_E}\left( {1 + {{\overline K }_E}} \right)}},\frac{{{{\overline \gamma  }_E}{m_B}\left( {{m_E} + {K_{E,i}}} \right)\left( {1 + {{\overline K }_B}} \right)}}{{{{\overline \gamma  }_B}{m_E}\left( {{m_B} + {K_{B,j}}} \right)\left( {1 + {{\overline K }_E}} \right)}},\frac{{{K_{E,i}}}}{{{m_E}}},\frac{{{{\overline \gamma  }_E}\left( {{m_E} + {K_{E,i}}} \right)}}{{{m_E}\left( {1 + {{\overline K }_E}} \right)}}} \right|} \right.\notag \\
		&\left. {\begin{array}{*{20}{c}}
			{\left( { - 1;1,1,1,1} \right)}\\
			{\left( { - 1;1,1,0,0} \right),\left( {0;0,0,1,0} \right),\left( {0;0,0,0,1} \right)}
			\end{array}\begin{array}{*{20}{c}}
			:\\
			:
			\end{array}\begin{array}{*{20}{c}}
			{\left( {{m_B},1} \right)}\\
			{\left( {0,1} \right)}
			\end{array}\begin{array}{*{20}{c}}
			;\\
			;
			\end{array}\begin{array}{*{20}{c}}
			{\left( {1 - {m_B},1} \right)}\\
			{\left( {0,1} \right)}
			\end{array}\begin{array}{*{20}{c}}
			;\\
			;
			\end{array}\begin{array}{*{20}{c}}
			{\left( {{m_E},1} \right)}\\
			{\left( {0,1} \right)}
			\end{array}\begin{array}{*{20}{c}}
			;\\
			;
			\end{array}\begin{array}{*{20}{c}}
			{\left( {1,1} \right),\left( {1,1} \right)}\\
			{\left( {1,1} \right)}
			\end{array}} \right],
		\end{align} }		
			\setcounter{equation}{\value{mytemp1}}
		\end{figure*}
		\setcounter{equation}{23}				
		{\small\begin{equation}\label{26}
		{Q_{\ell ,r}} \buildrel \Delta \over = \left\{ {\begin{array}{*{20}{c}}
			{{p_\ell }},\\
			{1 - {p_\ell }},
			\end{array}\begin{array}{*{20}{c}}
			{{\rm if}\quad r = 1}\\
			{{\rm if}\quad r = 2}
			\end{array}} \right., \quad \ell  \in \left\{ {B,E} \right\},
		\end{equation}}
which is the possibility of the first or the second ARS depending on.
\end{theorem}
\begin{IEEEproof}
	See Appendix \ref{AppendixA}.
\end{IEEEproof}
Substituting \eqref{24} and \eqref{25} into \eqref{21} and \eqref{22}, respectively, and substituting \eqref{21}, \eqref{22} and \eqref{23} into \eqref{17}, we can derive the exact expression of ASC.
		\begin{coro}\label{prop2}
			For integer values of $m$, ${I_1}$, ${I_2}$ and ${I_3}$ can be expressed as \eqref{21}, \eqref{22} and
			{\small \begin{align}\label{27}
			&{I_3} = {p_E}\sum\limits_{n = 0}^{{m_E} - 1} {\frac{{{B_{n,E,1}}}}{{\Gamma \left( {{m_E} - n} \right)}}G_{3,2}^{{\rm{1,3}}}\left( {{{\overline \rho  }_{E,1}}\left| {\begin{array}{*{20}{c}}
						{ - {m_E} + n + 1,1,1}\\
						{1,0}
						\end{array}} \right.} \right)}  \notag\\
			&+ \!\left( {1\! - \!{p_E}} \right)\!\!\sum\limits_{n = 0}^{{m_E} - 1} \!\!{\frac{{{B_{n,E,2}}}}{{\Gamma \left( {{m_E} - n} \right)}}G_{3,2}^{{\rm{1,3}}}\left( {{{\overline \rho  }_{E,2}}\left| \!\!\!{\begin{array}{*{20}{c}}
						{ - {m_E} + n + 1,1,1}\\
						{1,0}
						\end{array}} \right.} \right)}  ,
			\end{align}}
			where  ${R_{i,j}}$  and ${T_{i,j}}$ are derived as
			\eqref{28} and \eqref{29} at the top of
			the next page,
			\newcounter{mytemp2}
			\begin{figure*}[t]
				\normalsize
				\setcounter{mytemp2}{\value{equation}}
				\setcounter{equation}{25}
			{\small 	\begin{align}\label{28}
				&{R_{i,j}}\! \buildrel \Delta \over = {Q_{B,i}}{Q_{E,j}}\sum\limits_{l = 0}^{{m_E} - 1} {\sum\limits_{n = 0}^{{m_B} - 1} {\frac{{{B_{l,E,j}}{B_{n,B,i}}}}{{\Gamma \left( {{m_B} - n} \right)}}} } G_{3,2}^{{\rm{1,3}}}\left( {{{\overline \rho  }_{B,i}}\left| {\begin{array}{*{20}{c}}
						{1 - {m_B} + n,1,1}\\
						{1,0}
						\end{array}} \right.} \right) \notag\\
				&- {Q_{B,i}}{Q_{E,j}}\!\!\sum\limits_{l = 0}^{{m_E} - 1} {\sum\limits_{n = 0}^{{m_B} - 1} {\sum\limits_{k = 0}^{{m_E} - l - 1}\!\! {\frac{{{B_{l,E,j}}{B_{n,B,i}}\overline \rho  _{B,i}^k\overline \rho  _{E,j}^{{m_B} - n}}}{{\Gamma \left( {{m_B} - n} \right)k!{{\left( {{{\overline \rho  }_{E,j}} + {{\overline \rho  }_{B,i}}} \right)}^{k + {m_B} - n}}}}} } } G_{3,2}^{{\rm{1,3}}}\!\!\left( {\frac{{{{\overline \rho  }_{B,i}}{{\overline \rho  }_{E,j}}}}{{{{\overline \rho  }_{B,i}} + {{\overline \rho  }_{E,j}}}}\left| {\begin{array}{*{20}{c}}
						{ - k - {m_B} + n + 1,1,1}\\
						{1,0}
						\end{array}} \right.} \!\!\!\right),
				\end{align}}
				\hrulefill
				{\small \begin{align}\label{29}
					&	{T_{i,j}} \! \buildrel \Delta \over = {Q_{B,i}}{Q_{E,j}}\sum\limits_{l = 0}^{{m_B} - 1} {\sum\limits_{n = 0}^{{m_E} - 1} {{B_{l,B,i}}{B_{n,E,j}}\frac{1}{{\Gamma \left( {{m_E} - n} \right)}}} } G_{3,2}^{{\rm{1,3}}}\left( {{{\overline \rho  }_{E,j}}\left| {\begin{array}{*{20}{c}}
							{1 - {m_E} + n,1,1}\\
							{1,0}
							\end{array}} \right.} \right)\notag\\
					&- {Q_{B,i}}{Q_{E,j}}\!\!\sum\limits_{l = 0}^{{m_B} - 1} {\sum\limits_{n = 0}^{{m_E} - 1} {\sum\limits_{k = 0}^{{m_B} - l - 1}\!\!\! {\frac{{{B_{l,B,i}}{B_{n,E,j}}\overline \rho  _{E,j}^k\overline \rho  _{B,i}^{{m_E} - n}}}{{\Gamma \left( {{m_E} - n} \right)k!{{\left( {{{\overline \rho  }_{B,i}} + {{\overline \rho  }_{E,j}}} \right)}^{k + {m_E} - n}}}}} } } G_{3,2}^{{\rm{1,3}}}\!\!\left( {\frac{{{{\overline \rho  }_{B,i}}{{\overline \rho  }_{E,j}}}}{{{{\overline \rho  }_{B,i}} + {{\overline \rho  }_{E,j}}}}\left| {\begin{array}{*{20}{c}}
							{ - k - {m_E} + n + 1,1,1}\\
							{1,0}
							\end{array}} \right.} \!\!\!\!\right),
					\end{align}}
				\hrulefill
				\setcounter{equation}{\value{mytemp2}}
			\end{figure*}
			\setcounter{equation}{27}
			 $ G_{p,q}^{m,n}\left( \cdot \right) $ is the Meijer's $G$-function \cite[Eq. (9.301)]{gradshteyn2007}, $ \Gamma \left( \cdot \right) $ is the gamma function \cite[Eq. (8.310.1)]{gradshteyn2007}, and
		{\small	\begin{align}\label{30}
			{\overline \rho  _{\ell ,r}} \buildrel \Delta \over = \frac{{{K_{\ell ,r}} + {m_\ell }}}{{{m_\ell }\left( {1 + {{\overline K }_\ell }} \right)}}{\overline \gamma  _\ell },
			\end{align}
			\begin{align}\label{31}
			{B_{n,\ell ,r}}\! \buildrel \Delta \over =\! &\left( {\begin{array}{*{20}{c}}
				{{m_\ell }\! -\! 1}\\
				n
				\end{array}} \right)\!\!{\left( {\frac{{{m_\ell }}}{{{K_{\ell ,r}} + {m_\ell }}}} \right)^n}\!\!{\left( {\frac{{{K_{\ell ,r}}}}{{{K_{\ell ,r}} + {m_\ell }}}} \right)^{{m_\ell } - n - 1}}, \notag \\
			&\forall 0 \le n \le {m_\ell } - 1,\ell  \in \left\{ {B,E} \right\},r \in \left\{ {1,2} \right\}.
			\end{align}}
		\end{coro}
		\begin{IEEEproof}
			See Appendix \ref{AppendixB}.
		\end{IEEEproof}	

Substituting \eqref{28} and \eqref{29} into \eqref{21} and \eqref{22}, and substituting \eqref{21}, \eqref{22} and \eqref{27} into \eqref{17}, we derive the integer approximated ASC.
\subsection{Approximate ASC for High SNRs}
In order to explicitly examine the performance in the high-SNR regime, we derive the asymptotic ASC for the case of ${\overline \gamma  _B} \to \infty $. The asymptotic ASC expression for high SNRs can be expressed as
{\small\begin{align}\label{Csappr}
	{\overline C _{s,appr}} = {I_{1,appr}} + {I_{2,appr}}-{I_{\rm{3}}}.
\end{align}}

\begin{prop}\label{prop111}
	To obtain the asymptotic expression of the ASC for high SNRs,
	the asymptotic expressions for ${I_1}$ and ${I_2}$ are derived as
{\small	\begin{equation}\label{I1_appr}
	{I_{1,{\rm appr}}} \simeq {R_{1,1,{\rm appr}}} + {R_{1,2,{\rm appr}}} + {R_{2,1,{\rm appr}}} + {R_{2,2,{\rm appr}}},
	\end{equation}
	\begin{equation}\label{I2_appr}
	{I_{2,{\rm appr}}} \simeq {T_{1,1,{\rm appr}}} + {T_{1,2,{\rm appr}}} + {T_{2,1,{\rm appr}}} + {T_{2,2,{\rm appr}}},
	\end{equation}}
	where
{\small	\begin{align}\label{R_appr}
&{R_{i,j,appr}}\! \buildrel \Delta \over =\! {Q_{\!B,i}}{Q_{\!E,j}}\!\!\left(\!\! {\ln \!\!\left(\!\! {\frac{{{{\overline \gamma  }_B}\!\left( {{m_B}\! +\! {K_{1,B}}} \right)}}{{{m_B}\left( {1 + {{\overline K }_B}}\! \right)}}} \!\!\right)\! \!-\! C }\!\! \right)- \frac{{{Q_{B,i}}{Q_{E,j}}}}{{\Gamma \left( {1 - {m_B}} \right)}}
\notag\\
&\times\!\!\!
{\left(\!\! {\frac{{{m_B}}}{{{m_B}\!\! +\! \!{K_{B,i}}}}} \!\!\right)^{\!\!{m_B} \!- \!1}}\!\!\!\!\!\! G_{1,0;0,1;2,3}^{0,1;1,0;1,2}\!\!\left(\!\!\!\!\!{\left. {\begin{array}{*{20}{c}}
		1\\
		-
		\end{array}}\!\!\! \right|\!\!\!\!\!\left. {\begin{array}{*{20}{c}}
		- \\
		{1 \!-\! {m_B}}
		\end{array}} \!\!\!\!\right|\!\!\!\!\left. {\begin{array}{*{20}{c}}
		{0,0}\\
		{0, - \!1, - \!1}
		\end{array}}\!\!\!\! \right|\frac{{{m_B}}}{{{K_{B,i}}}}\!,\! -\! 1} \!\!\right)\!\!,
	\end{align}}
	{\small\begin{align}\label{T_appr}
	&{T_{i,j,appr}} \buildrel \Delta \over = \frac{{{Q_{E,i}}{Q_{B,j}}{{\overline \gamma  }_E}}}{{\Gamma \!\left( {1 \!-\! {m_E}} \right)}{{\overline \gamma  }_B}}
	 {\left(\! {\frac{{{m_B}}}{{{m_B} \!+\! {K_{B,j}}}}}\! \right)^{\!{m_B}}}\!{\left(\! {\frac{{{m_E}}}{{{m_E} \!+\! {K_{E,i}}}}} \!\right)^{\!{m_E} - 2}}\notag\\
	&\times \!\!
		\frac{{ {1 \!+\! {{\overline K }_B}}}}{{ {1\! +\! {{\overline K }_E}} }}
	G_{1,0;1,2;2,2}^{0,1;1,1;1,2}\left(\!\!\!\!\! {\left. {\begin{array}{*{20}{c}}
			2\\
			-
			\end{array}}\!\!\!\right|\!\!\!\!\!\left. {\begin{array}{*{20}{c}}
			{{m_E}}\\
			{0,0}
			\end{array}}\!\!\! \right|\!\!\!\!\left. {\begin{array}{*{20}{c}}
			{1,1}\\
			{1,0}
			\end{array}}\!\!\! \right|\!\!\frac{{{K_{E,i}}}}{{{m_E}}}\!,\!\frac{{{{\overline \gamma  }_E}\!\left( {{m_E}\! +\! {K_{E,i}}} \right)}}{{{m_E}\left( {1\! +\! {{\overline K }_E}} \right)}}} \!\!\right)\!,
	\end{align}}
	where $ C $ is an Euler's constant \cite[Eq. (8.367.1)]{gradshteyn2007} and  $ G_{{p_1},{q_1};{p_2},{q_2};{p_3},{q_3}}^{{m_1},{n_1};{m_2},{n_2};{m_3},{n_3}}\left(  \cdot  \right) $ is the extended generalized bivariate Meijer's $ G $-function \cite{shah1973generalizations}.
\end{prop}
\begin{IEEEproof}
		See Appendix \ref{AppendixCNEW}.
\end{IEEEproof}

Subsequently, substituting \eqref{R_appr} and \eqref{T_appr} into \eqref{I1_appr} and \eqref{I2_appr}, respectively, then into \eqref{Csappr}, we derive the asymptotic expression of the ASC for high SNRs.	
	
The high-SNR slope is a key performance indicator for ASC at high SNRs and is given by \cite[Eq. (10)]{lozano2005high}
		\begin{align}\label{Sinf}
		{S_\infty } = \mathop {\lim }\limits_{{{\overline \gamma  }_B} \to \infty } \frac{{{{\overline C }_{s,appr}}}}{{{{\log }_2}\left( {{{\overline \gamma  }_B}} \right)}}.
		\end{align}
Substituting \eqref{Csappr} into \eqref{Sinf}, we have
{\small\begin{align}
{S_\infty }&\!= \!\!\mathop {\lim }\limits_{{{\overline \gamma  }_B} \to \infty }\! \frac{{{I_{1,appr}}}}{{{{\log }_2}\left( {{{\overline \gamma  }_B}} \right)}}\! +\!\! \!\mathop {\lim }\limits_{{{\overline \gamma  }_B} \to \infty }\! \frac{{{I_{2,appr}}}}{{{{\log }_2}\left( {{{\overline \gamma  }_B}} \right)}}\! -\!\!\! \mathop {\lim }\limits_{{{\overline \gamma  }_B} \to \infty } \!\frac{{{I_3}}}{{{{\log }_2}\left( {{{\overline \gamma  }_B}} \right)}}\notag\\
&= {S_{\infty ,1}} + {S_{\infty ,2}} - {S_{\infty ,3}}
\end{align}}
With the help of  $ \mathop {\lim }\limits_{x \to \infty } \frac{{\ln \left( x \right)}}{{{{\log }_2}\left( x \right)}} = \ln \left( 2 \right) $, we derive
$ \mathop {\lim }\limits_{{{\overline \gamma  }_B} \to \infty } \frac{{{R_{i,j,appr}}}}{{{{\log }_2}\left( {{{\overline \gamma  }_B}} \right)}} = {Q_{B,i}}{Q_{E,j}}\ln \left( 2 \right). $ Substituting \eqref{I1_appr} into $ {S_{\infty ,1}}=\mathop {\lim }\limits_{{{\overline \gamma  }_B} \to \infty } \frac{{{I_{1,appr}}}}{{{{\log }_2}\left( {{{\overline \gamma  }_B}} \right)}} $, we derive ${S_{\infty ,1}} = \ln \left( 2 \right).$
With the help of $\mathop {\lim }\limits_{x \to \infty } \frac{1}{{x{{\log }_2}\left( x \right)}} = 0$, we derive ${S_{\infty ,2}} = 0. $
Since there is no ${{{\overline \gamma  }_B}}$ in \eqref{23}, ${S_{\infty ,3}}=0.$
Then \eqref{Sinf} is derived as
{\small\begin{align}\label{Sinf2}
{S_\infty } = \ln \left( 2 \right).
\end{align}}
\begin{rem}\label{rem1}
It is interesting to find from \eqref{Sinf2} that the shadowing parameter ${m_\ell }$, the Rician parameter  ${K_{\ell}}$ and
the possibility ${p_\ell }$ of both the main and wiretap channels are unrelated to the high-SNR slope.
For every 3dB increase in ${{{\overline \gamma  }_B}}$, ${\overline C }_{s,appr} $ increases by $\ln \left( 2 \right)$, when $ {{{\overline \gamma  }_B} \to \infty } $.
\end{rem}
		\section{Secrecy Outage Probability}
		\subsection{Exact and Integer Approximate SOP}
 Mathematically, the SOP can be evaluated by \cite[Eq. (7)]{kong2018physical}
		{\small\begin{align}\label{32}
		{P_{\rm o}} = \int_0^\infty  {{F_B}\left( {{R_s}{\gamma _E} + {R_s} - 1} \right){f_E}\left( {{\gamma _E}} \right)d{\gamma _E}},
		\end{align}}
		where $ {R_s} = {2^{{R_t}}}$. We first provide the following Theorem.
		\begin{theorem}\label{prop3}
		The exact and integer approximate SOP over ARS fading channels can be expressed as
		{\small\begin{align}\label{Pout_37}
		{P_{\rm o}} = {W_{1,1}} + {W_{1,2}} + {W_{2,1}} + {W_{2,2}},
		\end{align}}
        where $ {W_{i,j}} $ is derived as \eqref{338} for real values of $ m $ and \eqref{344} for integer values of $ m $
        at the top of the next page, respectively.
        \newcounter{mytempeqncnt}
        \begin{figure*}[t]
        	\normalsize
        	\setcounter{mytempeqncnt}{\value{equation}}
        	\setcounter{equation}{39}
        	{\small \begin{align}\label{338}
        		& {W_{i,j}} \buildrel \Delta \over = {Q_{B,i}}{Q_{E,j}}\frac{{1 + {{\overline K }_B}}}{{{{\overline \gamma  }_B}\Gamma \left( {1 - {m_E}} \right)}}{\left( {\frac{{{m_B}}}{{{m_B} + {K_{B,i}}}}} \right)^{{m_B}}}\!\!\!{\left( {\frac{{{m_E}}}{{{m_E} + {K_{E,j}}}}} \right)^{{m_E} - 1}} \!\!\!\!\!\sum\limits_{{n_1} = {n_2} = 0}^\infty \!\!\! {\sum\limits_{{n_3} = 0}^{1 + {n_1} + {n_2}} \!\!{\left( {\frac{{{{\left( {{R_s} - 1} \right)}^{1 + {n_1} + {n_2}}}\left( {1 + {n_1} + {n_2}} \right)!}}{{{n_1}!{n_2}!{n_3}!\left( {1 + {n_1} + {n_2} - {n_3}} \right)!}}} \right.} }\notag\\
        		&\times\frac{{{{\left( {1 - {m_B}} \right)}_{{n_1}}}{{\left( {{m_B}} \right)}_{{n_2}}}}}{{{{\left( 2 \right)}_{{n_1} + {n_2}}}}} \left. {{{\left( { - \frac{{1 + {{\overline K }_B}}}{{{{\overline \gamma  }_B}}}} \right)}^{{n_1}}}\!\!{{\left( { - \frac{{{m_B}\left( {1 + {{\overline K }_B}} \right)}}{{{{\overline \gamma  }_B}\left( {{m_B} + {K_{B,i}}} \right)}}} \right)}^{{n_2}}}\!\!{{\left( {\frac{{{R_s}{{\overline \gamma  }_E}\left( {{m_E} + {K_{E,j}}} \right)}}{{\left( {{R_s} - 1} \right)\left( {1 + {{\overline K }_E}} \right){m_E}}}} \right)}^{{n_3}}}\!\!\!G_{2,2}^{1,2}\left( {\frac{{{K_{E,j}}}}{{{m_E}}}\left| {\begin{array}{*{20}{c}}
        					{{m_E}, - {n_3}}\\
        					{0,0}
        					\end{array}} \!\!\right.} \right)}\!\! \right),
        		\end{align} }
        	\hrulefill
        	{\small \begin{align}\label{344}
        		{W_{\!i,j}} \!\buildrel \Delta \over =\! {Q_{\!B\!,i}}{Q_{\!E\!,j}}\!\!\!\!\sum\limits_{l = 0}^{{m_{\!B}}\! -\! 1} {\sum\limits_{n = 0}^{{m_{\!E}} \!-\! 1}\!\! {{B_{l,B,i}}{B_{n,E,j}}} }\! \left(\! {1 \!-\!\!\!\sum\limits_{k = 0}^{{m_{\!B}} \!-\! l\! -\! 1}\!\! {\sum\limits_{t = 0}^k \!\!{\left( \!{\frac{{{{\left( {{R_s} \!-\! 1} \right)}^t}\!R_s^{k \!-\! t}}}{{\bar \rho _{B,i}^k\bar \rho _{E,j}^{{m_E} - n}}}\!\!\left. {\frac{{\Gamma\! \left( {{m_{\!E}}\! -\! n\! +\! k \!-\! t} \right)}}{{t!\left( {k\! -\! t} \right)!\Gamma ({m_{\!E}} \!-\! n)}}\!\exp\!\! \left(\!\! { - \frac{{{R_s} \!-\! 1}}{{{{\bar \rho }_{B,i}}}}} \right)\!\!{{\left( \!{\frac{1}{{{{\bar \rho }_{E,j}}}}{\rm{ + }}\frac{{{R_s}}}{{{{\bar \rho }_{B,i}}}}} \!\right)}^{ \!- {m_E} +\! n -\! k +\! t}}} \right)} \right.} } } \!\!\!\right),
        		\end{align} }
        	\setcounter{equation}{\value{mytempeqncnt}}
        	\hrulefill
        \end{figure*}
        \setcounter{equation}{41}			
		\end{theorem}
		\begin{IEEEproof}
			See Appendix \ref{AppendixC}.
		\end{IEEEproof}
		
		\subsection{Approximate SOP for High SNRs}
		In order to explicitly show the physical insights, we derive the asymptotic SOP for ${\overline \gamma  _B} \to \infty $.
		\begin{prop}
			The asymptotic SOP for high SNRs is
			\begin{align}\label{SOPappr}
			{P_{{\rm o},{\rm appr}}} \!\simeq\! {W_{1,1,{\rm appr}}} \!+\! {W_{1,2,{\rm appr}}} \!+\! {W_{2,1,{\rm appr}}} \!+\! {W_{2,2,{\rm appr}}},
			\end{align}		
			where  $ W_{i,j,{\rm appr}} $ is derived as \eqref{W_appr} at the top of the next page		
		 and $N_L$ is a positive integer.
		\newcounter{mytempeqncnt1}
		\begin{figure*}[t]
			\normalsize
			\setcounter{mytempeqncnt1}{\value{equation}}
			\setcounter{equation}{42}
			{\small \begin{align}\label{W_appr}
				&{{W}_{i,j,{\rm appr}}}\! \buildrel \Delta \over =  {Q_{B,i}}{Q_{E,j}}\frac{{1 + {{\overline K }_B}}}{{{{\overline \gamma  }_B}\Gamma \left(\! {1 - {m_E}} \!\right)}}{\left( {\frac{{{m_B}}}{{{m_B} + {K_{B,i}}}}} \right)^{{m_B}}}\!\!\!{\left(\!\! {\frac{{{m_E}}}{{{m_E} + {K_{E,j}}}}} \!\!\right)^{{m_E} - 1}}\!\!\!\!\! \sum\limits_{{n_1} = {n_2} = 0}^{N_L} \!\!\!{\sum\limits_{{n_3} = 0}^{1 + {n_1} + {n_2}} \!\!\!{\left( {\frac{{{{\left( {{R_s} - 1} \right)}^{1 + {n_1} + {n_2}}}\left( {1 + {n_1} + {n_2}} \right)!}}{{{n_1}!{n_2}!{n_3}!\left( {1 + {n_1} + {n_2} - {n_3}} \right)!}}} \right.} }\notag\\
				&\times\!\! \frac{{{{\left( \!{1 - {m_B}} \!\right)}_{{n_1}}}{{\left( {{m_B}} \right)}_{{n_2}}}}}{{{{\left( 2 \right)}_{{n_1} + {n_2}}}}}\left. {{{\left(\!\! { - \frac{{1 + {{\overline K }_B}}}{{{{\overline \gamma  }_B}}}} \!\right)}^{{n_1}}}{{\left( { - \frac{{{m_B}\left( {1 + {{\overline K }_B}} \right)}}{{{{\overline \gamma  }_B}\left( {{m_B} + {K_{B,i}}} \right)}}} \right)}^{{n_2}}}\!\!{{\left( {\frac{{{R_s}{{\overline \gamma  }_E}\left( {{m_E} + {K_{E,j}}} \right)}}{{\left( {{R_s} - 1} \right)\left( {1 + {{\overline K }_E}} \right){m_E}}}} \right)}^{{n_3}}}\!\!G_{2,2}^{1,2}\left( {\frac{{{K_{E,j}}}}{{{m_E}}}\left| {\begin{array}{*{20}{c}}
							{{m_E}, - {n_3}}\\
							{0,0}
							\end{array}} \right.} \!\!\right)}\!\! \right),
				\end{align}}
			\hrulefill
					{\small \begin{align}\label{ss3}
						&{{\hat W}_{i,j}}\buildrel \Delta \over = {Q_{B,i}}{Q_{E,j}}\frac{{1 + {{\overline K }_B}}}{{{{\overline \gamma  }_B}\Gamma \left( {1 - {m_E}} \right)}}{\left( {\frac{{{m_B}}}{{{m_B} + {K_{B,i}}}}} \right)^{{m_B}}}\!\!{\left( {\frac{{{m_E}}}{{{m_E} + {K_{E,j}}}}} \right)^{{m_E} - 1}} \!\!\!\!\sum\limits_{{n_1} = {n_2} = 0}^{N_l} \!\!\!{\sum\limits_{{n_3} = 0}^{1 + {n_1} + {n_2}} \!\!\!{\left( {\frac{{{{\left( {{R_s} - 1} \right)}^{1 + {n_1} + {n_2}}}\left( {1 + {n_1} + {n_2}} \right)!}}{{{n_1}!{n_2}!{n_3}!\left( {1 + {n_1} + {n_2} - {n_3}} \right)!}}} \right.} }\notag\\
						&\times \!\!\frac{{{{\left( {1 - {m_B}} \right)}_{{n_1}}}\!{{\left( {{m_B}} \right)}_{{n_2}}}}}{{{{\left( 2 \right)}_{{n_1} + {n_2}}}}}\left. {{{\left( { - \frac{{1 + {{\overline K }_B}}}{{{{\overline \gamma  }_B}}}} \right)}^{{n_1}}}\!\!\!{{\left( { - \frac{{{m_B}\left( {1 + {{\overline K }_B}} \right)}}{{{{\overline \gamma  }_B}\left( {{m_B} + {K_{B,i}}} \right)}}} \right)}^{{n_2}}}\!\!\!{{\left( {\frac{{{R_s}{{\overline \gamma  }_E}\left( {{m_E} + {K_{E,j}}} \right)}}{{\left( {{R_s} - 1} \right)\left( {1 + {{\overline K }_E}} \right){m_E}}}} \right)}^{{n_3}}}\!\!\!G_{2,2}^{1,2}\left( {\frac{{{K_{E,j}}}}{{{m_E}}}\left| {\begin{array}{*{20}{c}}
									{{m_E}, - {n_3}}\\
									{0,0}
									\end{array}} \right.} \!\!\right)} \!\!\right).
									\tag{49}
						\end{align}}
			
			\setcounter{equation}{\value{mytempeqncnt1}}
			\hrulefill
		\end{figure*}
		\setcounter{equation}{43}
	\end{prop}
	
		\begin{IEEEproof}	
					The SOP expression can be expressed as
	{\small 	\begin{align}
		&{W_{i,j,{\rm{appr}}}} \buildrel \Delta \over =  {Q_{B,i}}{Q_{E,j}}{\left( {\frac{{{m_E}}}{{{m_E} + {K_{E,j}}}}} \right)^{{m_E} - 1}}{\left( {\frac{{{m_B}}}{{{m_B} + {K_{B,i}}}}} \right)^{{m_B}}}\!\notag\\
		&\times\frac{{1 + {{\bar K}_B}}}{{{{\bar \gamma }_B}\Gamma \left( {1 \!-\! {m_E}} \right)}}\!\left( {\sum\limits_{{n_1} = {n_2} = 0}^{30} \!\!\!{\sum\limits_{{n_3} = 0}^{1 + {n_1} + {n_2}}\!\! \Xi  }  +\!\!\! \sum\limits_{{n_1} = {n_2} = {\rm{31}}}^\infty  \!\!\!{\sum\limits_{{n_3} = 0}^{1 + {n_1} + {n_2}} \!\!\Xi  } } \right),
		\end{align}}
where
		{\small \begin{align}
		&\Xi  \buildrel \Delta \over =  \frac{{{{\left( {{R_s} - 1} \right)}^{1 + {n_1} + {n_2}}}\left( {1 + {n_1} + {n_2}} \right)!}}{{{n_1}!{n_2}!{n_3}!\left( {1 + {n_1} + {n_2} - {n_3}} \right)!}}\frac{{{{\left( {1 - {m_B}} \right)}_{{n_1}}}{{\left( {{m_B}} \right)}_{{n_2}}}}}{{{{\left( 2 \right)}_{{n_1} + {n_2}}}}}
		\notag\\
		&\times
		{\left( { - \frac{{1 + {{\bar K}_B}}}{{{{\bar \gamma }_B}}}} \right)^{{n_1}}} {\left( { - \frac{{{m_B}\left( {1 + {{\bar K}_B}} \right)}}{{{{\bar \gamma }_B}\left( {{m_B} + {K_{B,i}}} \right)}}} \right)^{{n_2}}}\notag\\
		&\times{\left( {\frac{{{R_s}{{\bar \gamma }_E}\left( {{m_E} + {K_{E,j}}} \right)}}{{\left( {{R_s} - 1} \right)\left( {1 + {{\bar K}_E}} \right){m_E}}}} \right)^{{n_3}}}\!\!\!G_{2,2}^{1,2}\!\!\left( {\frac{{{K_{E,j}}}}{{{m_E}}}\!\left| \!\!{\begin{array}{*{20}{c}}
				{{m_E}, - {n_3}}\\
				{0,0}
				\end{array}} \right.}\!\!\! \right).
		\end{align}}
		It can be shown that $\sum\limits_{{n_1} = {n_2} = {N_L+1}}^\infty  {\sum\limits_{{n_3} = 0}^{1 + {n_1} + {n_2}} \Xi  }  \to 0 $ as $ {{\bar \gamma }_B} \to \infty  $. We obtain the approximate $ W_{i,j} $ as \eqref{W_appr} to complete the proof.
		\end{IEEEproof}
		
		Based on \eqref{SOPappr}, we derive the SDO which is a key performance indicator for asymptotic SOP.
	    The SDO is given by \cite{zheng2003diversity}
	    {\small\begin{align}\label{SDO}
	    {G_d} =  - \mathop {\lim }\limits_{{{\overline \gamma  }_B} \to \infty } \frac{{\log \left( {{P_{\rm o,appr}}} \right)}}{{\log \left( {{{\overline \gamma  }_B}} \right)}}.
	    \end{align}}
		Substituting \eqref{SOPappr} into \eqref{SDO} and with the help  of $ \sum\limits_{n = {n_0}}^\infty  {\frac{1}{{{x^n}}}}  \to \frac{1}{{{x^{{n_0}}}}} $ as $x \to \infty  $, we derive
		{\small \begin{align}\label{Gd=1}
	{G_d} = 1.
		\end{align}}
		\begin{rem}\label{rem2}
			From \eqref{Gd=1}, an interesting result can be observed that the SDO is  a constant. It demonstrates that SDO
			is independent of the shadowing parameter ${m_\ell }$,
			the Rician parameter  ${K_{\ell}}$ and the possibility ${p_\ell }$ of both the main and wiretap channels, respectively.
		\end{rem}
		\subsection{Truncation Error}
	    There are an infinite number of summation terms in \eqref{338}. It is difficult to calculate the results of SOP for real values of $m$.
	    Fortunately, the accurate results can be derived by summing only up to a small number of terms. Therefore, it is necessary to calculate the truncation error.
	
		By truncating \eqref{Pout_37} up to the first $ {N_l} $ terms, we have the approximate outage
		{\small\begin{align}
		{{\hat P}_{\rm o}} = {{\hat W}_{1,1}} + {{\hat W}_{1,2}} + {{\hat W}_{2,1}} + {{\hat W}_{2,2}},
		\end{align}}
		where $ {{\hat W}_{i,j}} $ is derived as \eqref{ss3} at the top of this page.

	\setcounter{equation}{49}
		The truncation error of the area under $ P_{o} $ with respect to the first $ N_l $ terms is given by
		{\small\begin{align}
		\varepsilon \left( {{N_l}} \right) \buildrel \Delta \over = \left| {{P_{{\rm{o}}}} - {{\hat P}_{{\rm{o}}}}} \right|.
		\end{align}}

		In order to demonstrate the convergence of the series in \eqref{Pout_37} for real values of $ m $, Table \ref{table1} depicts the required truncation terms $ N_l $ for different channel parameters. It is clear to see that the number of required truncation terms increases as $ {K_{B,i}}$, ${K_{E,j}} $, $ {\overline \gamma  _E} $ and $ m_B $ increase, and
		decreases as $ {\overline \gamma  _B} $ increases. Furthermore, Eq. \eqref{Pout_37} converges to a limit value with a small number of terms.			
		{  \begin{table}[t]
				\caption{\label{table1}Required Terms of The Truncation Error ($ \varepsilon  < {10^{ - 6}} $) for Different Parameters with $ {m_E} = 0.5$, ${R_t} = 0.5 $}.
\footnotesize
				\centering
				\begin{tabular}{|c|c|c|}
					\toprule
					\hline
					Parameters & $ {N_l} $ & $\varepsilon $ \\
					\hline
					\makecell[c]{$ {K_{B,{\rm{1}}}}{\rm{ = 30}}$ , ${K_{B,{\rm{2}}}}{\rm{ = 10}}$ ,\\ ${K_{E,{\rm{1}}}}{\rm{ = 30}}$ , ${K_{E,{\rm{2}}}}{\rm{ = 10}}$, ${m_B} = 10 $ ,\\ ${\overline \gamma  _B} = 30$  ${\rm dB}$ , ${\overline \gamma  _E} = 10 $  ${\rm dB}$ } & 33 & $ 7.32 \times {10^{ - 7}} $\\
					\hline
					\makecell[c]{$ {K_{B,{\rm{1}}}}{\rm{ = 30}}$ , ${K_{B,{\rm{2}}}}{\rm{ = 10}}$ ,\\ ${K_{E,{\rm{1}}}}{\rm{ = 60}}$ , ${K_{E,{\rm{2}}}}{\rm{ = 20}}$, ${m_B} = 10 $ ,\\ ${\overline \gamma  _B} = 30$  ${\rm dB}$ , ${\overline \gamma  _E} = 10 $  ${\rm dB}$ } & 35  &  $ 3.99 \times {10^{ - 7}} $ \\
					\hline
					\makecell[c]{$ {K_{B,{\rm{1}}}}{\rm{ = 60}}$ , ${K_{B,{\rm{2}}}}{\rm{ = 20}}$ ,\\ ${K_{E,{\rm{1}}}}{\rm{ = 30}}$ , ${K_{E,{\rm{2}}}}{\rm{ = 10}}$, ${m_B} = 10 $ ,\\ ${\overline \gamma  _B} = 30$  ${\rm dB}$ , ${\overline \gamma  _E} = 10 $  ${\rm dB}$ } & 56 &  $ 4.28 \times {10^{ - 7}} $  \\
					\hline
					\makecell[c]{$ {K_{B,{\rm{1}}}}{\rm{ = 30}}$ , ${K_{B,{\rm{2}}}}{\rm{ = 10}}$ ,\\ ${K_{E,{\rm{1}}}}{\rm{ = 30}}$ , ${K_{E,{\rm{2}}}}{\rm{ = 10}}$ , ${m_B} = 12 $,\\ ${\overline \gamma  _B} = 30$  ${\rm dB}$ , ${\overline \gamma  _E} = 10 $  ${\rm dB}$ } & 46 & $ 6.53 \times {10^{ - 7}} $  \\
					\hline
					\makecell[c]{$ {K_{B,{\rm{1}}}}{\rm{ = 30}}$ , ${K_{B,{\rm{2}}}}{\rm{ = 10}}$ ,\\ ${K_{E,{\rm{1}}}}{\rm{ = 30}}$ , ${K_{E,{\rm{2}}}}{\rm{ = 10}}$, ${m_B} = 10 $ ,\\ ${\overline \gamma  _B} = 30$  ${\rm dB}$ , ${\overline \gamma  _E} = 8 $  ${\rm dB}$ } & 16 & $ 4.57 \times {10^{ - 7}} $ \\
					\hline
					\makecell[c]{$ {K_{B,{\rm{1}}}}{\rm{ = 60}}$ , ${K_{B,{\rm{2}}}}{\rm{ = 20}}$ ,\\ ${K_{E,{\rm{1}}}}{\rm{ = 30}}$ , ${K_{E,{\rm{2}}}}{\rm{ = 10}}$ , ${m_B} = 10 $,\\ ${\overline \gamma  _B} = 35$  ${\rm dB}$ , ${\overline \gamma  _E} = 10 $  ${\rm dB}$  }& 8 & $ 1.06 \times {10^{ - 7}} $ \\
					\hline
					\bottomrule
				\end{tabular}
			\end{table}}

		\section{Probability of Non-Zero Secrecy Capacity}
		The probability of non-zero secrecy capacity is defined as \cite[Eq. (11)]{kong2018physical}
		{\small\begin{align}\label{33}
		{P_{\rm nz}} = \int_0^\infty  {{F_E}\left( {{\gamma _B}} \right){f_B}\left( {{\gamma _B}} \right)d{\gamma _B}} .
		\end{align}}
		\begin{theorem}\label{prop4}
			The PNZ over ARS fading channels can be expressed as
			{\small\begin{align}\label{399}
			{P_{\rm nz}} = {D_{1,1}} + {D_{1,2}} + {D_{2,1}} + {D_{2,2}},
			\end{align}}
			where $ {D_{i,j}} $ is derived as \eqref{35} at the top of the next page			
			 for real values of $ m $
			 	\newcounter{mytempeqncnt11}
			 	\begin{figure*}[t]
			 		\normalsize
			 		\setcounter{mytempeqncnt11}{\value{equation}}
			 		\setcounter{equation}{52}
			 		{\small
			 				\begin{align}\label{35}
			 				&{D_{i,j}} \buildrel \Delta \over =
			 				\frac{{Q_{B,i}}{Q_{E,j}}}{{\Gamma \left( {1 - {m_B}} \right)\Gamma \left( {1 - {m_E}} \right)\Gamma \left( {{m_E}} \right)}}
			 				\frac{{{{\overline \gamma  }_B}\left( {1 + {{\overline K }_E}} \right)}}{{{{\overline \gamma  }_E}\left( {1 + {{\overline K }_B}} \right)}}{\left( {\frac{{{m_E}}}{{{m_E} + {K_{E,j}}}}} \right)^{{m_E}}}{\left( {\frac{{{m_B}}}{{{m_B} + {K_{B,i}}}}} \right)^{{m_B} - 2}}\notag\\
			 				& \times\!\! H_{1,2:1,1;1,1;1,1}^{0,1:1,1;1,1;1,1}\!\!\left[ {\frac{{{{\overline \gamma  }_{\!B}}\!\left( {{m_{\!B}}\! +\! {K_{\!B,i}}} \right)\!\left( {1 \!+ \!{{\overline K }_{\!E}}} \right)}}{{{{\overline \gamma  }_E}{m_B}\left( {1 + {{\overline K }_B}} \right)}}\!,\!\frac{{{{\overline \gamma  }_{\!B}}{m_{\!E}}\!\left( {{m_{\!B}} \!+ \!{K_{\!B,i}}} \right)\left( {1 \!+ \!{{\overline K }_{\!E}}} \right)}}{{{{\overline \gamma  }_{\!E}}{m_{\!B}}\!\left( {{m_{\!E}} \!+\! {K_{\!E,j}}} \right)\!\left( {1 \!+\! {{\overline K }_{\!B}}} \right)}}\!,\!\frac{{{K_{\!B,i}}}}{{{m_{\!B}}}}} \!\!\!\right.\left. {\left|\!\!\! {\begin{array}{*{20}{c}}
			 						{\left( { - \!1;\!1,\!1,\!1} \right)}\\
			 						{\left( { - \!1;\!1,\!1,\!0} \right),\left( {0;\!0,\!0,\!1} \right)}
			 						\end{array}\begin{array}{*{20}{c}}
			 						\!\!\!\!\!\!\!:\\
			 						\!\!\!\!\!\!\!:
			 						\end{array}\!\!\!\!\!\!\begin{array}{*{20}{c}}
			 						{\left( {{m_E},\!1} \right)\!\!:}\\
			 						{\left( {0,\!1} \right):}
			 						\end{array}\!\!\!\!\!\!\begin{array}{*{20}{c}}
			 						{\left( {1 \!-\! {m_E},\!1} \right)\!\!:}\\
			 						{\left( {0,1} \right):}
			 						\end{array}\!\!\!\!\!\!\begin{array}{*{20}{c}}
			 						{\left( {{m_B},\!1} \right)}\\
			 						{\left( {0,1} \right)}
			 						\end{array}} \right.}\!\!\!\! \right],
			 				\end{align} 
			 			}
			 		 	\hrulefill
			 		 		    {\small \begin{align}\label{pnz_appr}
			 		 		    	{D_{\!i\!,j\!,appr}} \!\!\buildrel \Delta \over =\!\! \frac{{{Q_{\!B,i}}{Q_{\!E,j}}}}{{\Gamma\!\! \left(\! {1\!\!-\!\! {m_{\!B}}} \!\right)\!\Gamma\!\! \left(\! {{m_{\!E}}}\! \right)}}\!{\left(\!\! {\frac{{{m_{\!E}}{{\overline \gamma  }_{\!B}}\!\left( {{m_{\!B}}\!\! +\!\! {K_{\!B\!,i}}} \right)\!\!\left( {1\! \!+\!\! {{\overline K }_{\!E}}} \right)}}{{{m_{\!B}}{{\overline \gamma  }_{\!E}}\!\left( {{m_{\!E}} \!\!+\!\! {K_{\!E\!,j}}} \right)\!\left( {1\!\! +\!\! {{\overline K }_{\!B}}} \right)}}}\!\! \right)^{\!\!\!{m_{\!E}}}}\!\!\!\!{\left(\! \!{\frac{{{m_{\!B}}}}{{{m_{\!B}}\!\! +\!\! {K_{\!B\!,i}}}}}\!\! \right)^{\!\!\!{m_{\!B}}\! -\! {\rm{1}}}}\!\!\!\! G_{\!0,1;2,1;2,1}^{\!1,0;1,1;1,1}\!\!\left(\!\!\!\!\!\! {\left. {\begin{array}{*{20}{c}}
			 		 		    			- \\
			 		 		    			{ - \!{m_{\!E}}}
			 		 		    			\end{array}} \!\!\!\!\right|\!\!\!\!\!\!\left. {\begin{array}{*{20}{c}}
			 		 		    			{1\!,\!1\!\! +\!\! {m_{\!E}}}\\
			 		 		    			{{m_E}}
			 		 		    			\end{array}}\!\!\!\! \right|\!\!\!\!\!\!\left. {\begin{array}{*{20}{c}}
			 		 		    			{1,1}\\
			 		 		    			{1\!\!-\!\! {m_{\!B}}}
			 		 		    			\end{array}} \!\!\!\!\right|\!\!\!\frac{{{{\overline \gamma  }_{\!E}}{m_{\!B}}\!\!\left( {{m_{\!E}}\!\! +\!\! {K_{\!E,j}}} \right)\!\!\left(\! {1\!\! +\!\! {{\overline K }_{\!B}}}\! \right)}}{{{{\overline \gamma  }_{\!B}}{m_{\!E}}\!\!\left(\! {{m_{\!B}} \!\!+\!\! {K_{\!B,i}}} \!\right)\!\!\left( {1\!\! +\!\! {{\overline K }_{\!E}}} \right)}}\!,\!\frac{{{m_{\!B}}}}{{{K_{\!B\!,i}}}}} \!\!\right)\!\!.
			 		 		    	\tag{56}
			 		 		    	\end{align}}			
			 		 		    \hrulefill 		 	
			 		\setcounter{equation}{\value{mytempeqncnt11}}
			 	\end{figure*}
			 	\setcounter{equation}{53}
			 	 and
			{\small \begin{align}\label{36}
				&{D_{i,j}} \buildrel \Delta \over = {Q_{B,i}}{Q_{E,j}}\sum\limits_{l = 0}^{{m_E} - 1} {\sum\limits_{n = 0}^{{m_B} - 1} {{B_{l,E,j}}{B_{n,B,i}}} } \notag\\
				&\times\!\!\left( \!{1\! - \!\!\!\!\!\!\!\sum\limits_{k = 0}^{{m_E} - l - 1} \!\!\!\!\! \frac{{\left( {k \!+\! {m_B} \!-\! n\! -\! 1} \right)!}}{{\bar \rho _{B,i}^{{m_B}\! -\! n}\bar \rho _{\!E\!,j}^k\Gamma\! \left( {{m_B}\! -\! n} \right)\!k!}}\!{{\left( \!{\frac{1}{{{{\bar \rho }_{B,i}}}}\! +\! \frac{1}{{{{\bar \rho }_{E,j}}}}}\! \right)}^{ \!\!\!- \!k\! -\! {m_B}\! +\! n}}}\! \right)\!\! ,
				\end{align} }				
 for integer values of $ m $, respectively.	
		\end{theorem}
		
		\begin{IEEEproof}
			See Appendix \ref{AppendixD}.
		\end{IEEEproof}

		In order to explicitly examine the performance in the high-SNR regime, we derive the asymptotic PNZ for ${\overline \gamma  _B} \to \infty $.
		\begin{prop}
	    The asymptotic expression of the PNZ for high SNRs can be expressed as
	    \begin{align}
	    {P_{{\rm nz},{\rm appr}}} \!\simeq {D_{1,1,{\rm appr}}}\! +\! {D_{1,2,{\rm appr}}} \!+\! {D_{2,1,{\rm appr}}}\! +\! {D_{2,2,{\rm appr}}},
	    \end{align}
	    where  $ D_{i,j,{\rm appr}} $   is derived as
	    	\eqref{pnz_appr} at the top of this page.
	
%

		\end{prop}
		\begin{IEEEproof}
				The asymptotic PNZ expression    for high SNRs can be obtained by computing the residues \cite{hu2019performance}. By considering the residue at the minimum pole on the right $ \left( {{r_1} = 1 - {m_E}} \right) $ in \eqref{35}, we finish the proof by obtaining the approximate expression of $ D_{i,j} $ as \eqref{pnz_appr}.
		\end{IEEEproof}
		\begin{rem}
			By considering the residue at the minimum pole on the right $ \left( {{r_2} = {m_E}} \right) $ in \eqref{pnz_appr} and after some algebraic manipulations, we can obtain that the asymptotic PNZ expression equals to the constant value one. 			
			It mathematically proves that the existence of of non-zero secrecy
			capacity is a certain event when the value of the main channel's SNR is large enough, i.e.,  $ {\overline \gamma  _B} \to \infty  $.
		\end{rem}

		\section{Numerical Results}
		
		In this section, our derived results are verified via Monte-Carlo simulations with $ {\rm{1}}{{\rm{0}}^{\rm{6}}} $ realizations. The parameters of main and wiretap channels are assumed to be independent and non-identically distributed random variables.
		
\begin{figure}[t]
				\centering
				\includegraphics[scale=0.5]{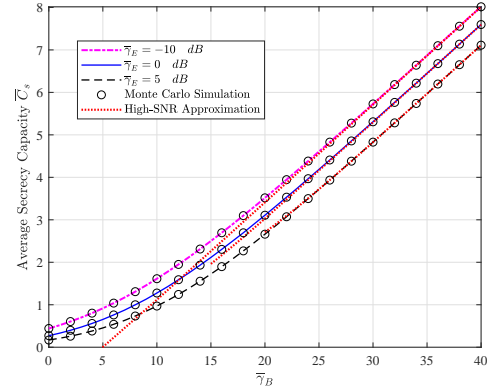}
				\caption{Average  Secrecy Capacity versus average SNR for $ {p_B} = {p_E} = 0.5 $, $ {m_B} = {m_E} = 0.5 $, $ {K_{B,1}} = {K_{E,1}} = 50/3 $, $ {K_{B,2}} = {K_{E,2}} = 10/3 $.}
				\label{ASC_1}
			\end{figure}

\begin{figure}[t]
				\centering
				\includegraphics[scale=0.5]{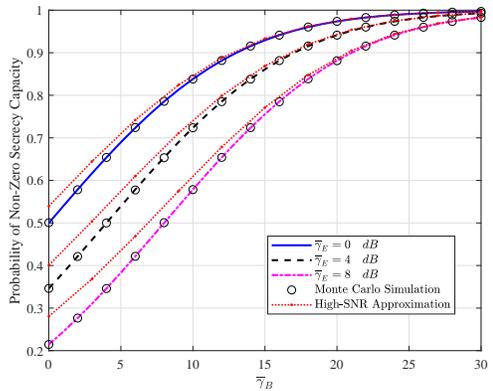}
				\caption{Probability of Non-Zero Secrecy Capacity versus average SNR for $ {p_B} = {p_E} = 0.5 $, $ {m_B} = {m_E} = 0.5 $, $ {K_{B,1}} = {K_{E,1}} = 50/3 $, $ {K_{B,2}} = {K_{E,2}} = 10/3 $.}
				\label{PNZ_1}
			\end{figure}

		Figures \ref{ASC_1} and \ref{PNZ_1} plot the ASC and PNZ curves versus $ {\overline \gamma  _B} $, for $ {p_B} = {p_E} = 0.5 $, $ {m_B} = {m_E} = 0.5 $, $ {K_{B,1}} = {K_{E,1}} = 50/3 $ and $ {K_{B,2}} = {K_{E,2}} = 10/3$. As expected, remarkable improvements can be achieved in the secrecy performance  as ${\overline \gamma _E} $ decreases, because of the deterioration in the TA-Eve wireless channel.
		From Fig. \ref{ASC_1}, one can observe that the high-SNR slope is a constant.
		For every 3dB increase in ${{{\overline \gamma  }_B}}$, ${\overline C }_{s,appr} $ increases by $\ln \left( 2 \right)$, when $ {{{\overline \gamma  }_B} \to \infty } $.
		The shadowing parameter ${m_\ell }$,
		the Rician parameter  ${K_{\ell}}$ and the possibility ${p_\ell }$ of both the main and wiretap channels have no impact on the high-SNR slope, which has been in Remark \ref{rem1}.
		In addition, the exact results match the Monte-Carlo simulations well and the asymptotic expressions match well the exact ones at the high-SNR regime, which proves their validity and versatility.
		
			
\begin{figure}[t]
			\centering
			\includegraphics[scale=0.5]{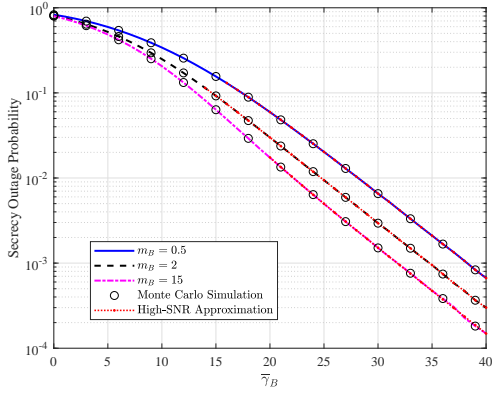}
			\caption{Secrecy Outage Probability versus average SNR for ${\overline \gamma  _E} = 4$  $\rm dB $, $ {p_B} = {p_E} = 0.5 $, $ {m_E} =0.5 $, $ {K_{B,1}} = {K_{E,1}} =50 $, $ {K_{B,2}} = {K_{E,2}} =10 $.}
			\label{SOP_2}
		\end{figure}

			Figure \ref{SOP_2} plots the SOP versus ${\overline \gamma  _B}$, for  ${\overline \gamma  _E} = 4$  $\rm dB $, $ {p_B} = {p_E} = 0.5 $, $ {m_E} = 0.5 $, $ {K_{B,1}} = {K_{E,1}} =50 $ and $ {K_{B,2}} = {K_{E,2}} =10 $. As it can be observed, larger values of $ {m_B} $ assures SOP with a lower probability. This is because of the fact that $m$ describes the level of fluctuation of the LoS ranging from 0.5 to $\infty $ and $m \to \infty$ means no fluctuation or constant LoS component. Besides, one can observe that SDO is a constant.  The shadowing parameter ${m_\ell }$,
			the Rician parameter  ${K_{\ell}}$ and the possibility ${p_\ell }$ of both the main and wiretap channels have no impact on SDO, which validates in Remark \ref{rem2}.  Moreover, analytical results also agree well with Monte-Carlo simulations.
		
\begin{figure}[t]
	\centering
	\includegraphics[scale=0.5]{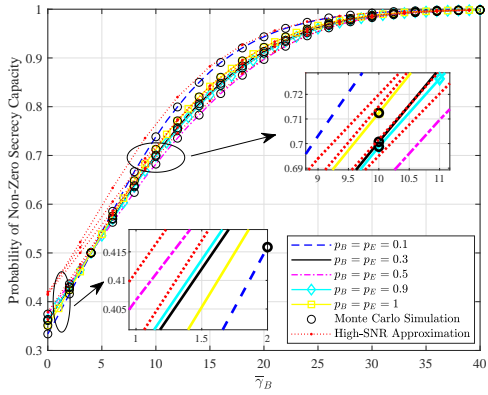}
	\caption{Probability of Non-Zero Secrecy Capacity versus average SNR for $ {\overline \gamma  _E} = 4$  $\rm dB$,  ${m_B} = {m_E} = 0.5$,  ${K_{B,1}} = {K_{E,1}} = 60$,  ${K_{B,2}} = {K_{E,2}} = 3 $.}
	\label{pnz_pi}
\end{figure}

Figure \ref{pnz_pi} plots the PNZ versus ${\overline \gamma  _B}$, for $ {\overline \gamma  _E} = 4$  $\rm dB$,  ${m_B} = {m_E} = 0.5$,  ${K_{B,1}} = {K_{E,1}} = 60$,  ${K_{B,2}} = {K_{E,2}} = 3 $. As it can be observed, when $ {\overline \gamma  _B} < 4$  $\rm dB $, PNZ increases first and then decreases after ${p_B} = {p_E} = 0.5 $, as $ {p_B} $ and $ {p_E} $ increases.
When $ {\overline \gamma  _B} > 4$  $\rm dB $, PNZ decreases first and then increases after ${p_B} = {p_E} = 0.5 $, as $ {p_B} $ and $ {p_E} $ increases.  Besides, all curves  intersect at $ {\overline \gamma  _B} = {\overline \gamma  _E} $, i.e., $ {\overline \gamma  _B} =4$ $\rm dB $.
When $ {p_B} \to 0.5$ and ${p_E} \to 0.5 $, remarkable improvements can be achieved in the secrecy performance  for $ {\overline \gamma  _B} < 4 $ $\rm dB$, but  for $ {\overline \gamma  _B} > 4 $ $\rm dB$, the secrecy performance degrades.
 Again, perfect agreement is observed between analytical results and Monte-Carlo simulations.

\begin{figure}[t]
				\centering
				\includegraphics[scale=0.5]{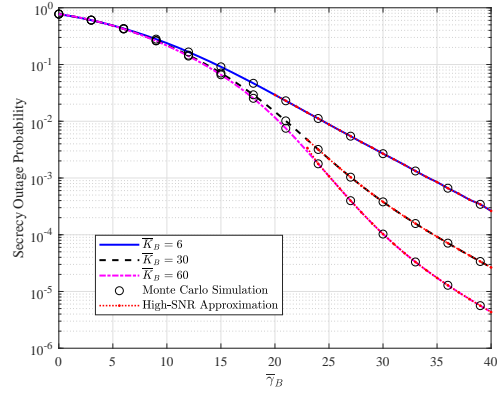}
				\caption{Secrecy Outage Probability versus average SNR for  ${\overline \gamma  _E} = 4$  $\rm dB $, $ {p_{\!B}}\! \!=\!\! {p_{\!E}}\!\! =\!\!\ 0.5 $, $ {m_{\!B}}\!\!=\!\!5$, ${m_{\!E}}\!\! =\!\! 0.5 $, $ {\overline K _{\!E}} \!\!=\!\! 6 $, $ {K_{\!B,1}}{\rm{/}}{K_{\!B,2}}\!\!=\!\!{K_{\!E,1}}{\rm{/}}{K_{\!E,2}}\!\!=\!\!5 $.}
				\label{SOP_3}
			\end{figure}

\begin{figure}[t]
	\centering
	\includegraphics[scale=0.5]{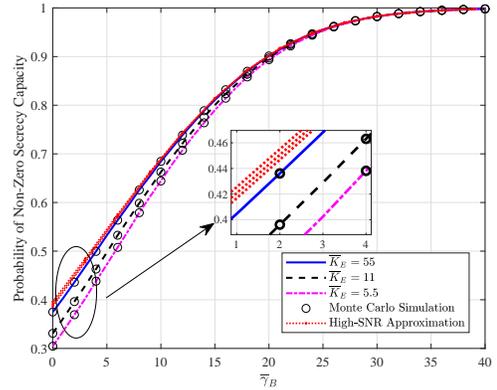}
	\caption{Probability of Non-Zero Secrecy Capacity versus average SNR for $ {\overline \gamma  _{\!E}}\! \!=\!\! 4$\!  $\rm  dB$, ${p_{\!B}}\!\! =\!\! {p_{\!E}}\! \!=\!\! 0.5$, ${m_{\!B}}\!\! =\!\! {m_{\!E}}\! \!=\!\! 0.5$, ${K_{\!B,1}}\! \!=\!\! 100$, ${K_{\!B,2}}\! \!=\!\! 10$,  $ {K_{\!E,1}}{\rm{/}}{K_{\!E,2}}\!\!=\!\!10 $.}
	\label{pnz_ki}
\end{figure}

		Figure \ref{SOP_3} illustrates the SOP versus ${\overline \gamma  _B}$, for  ${\overline \gamma  _E} = 4$  $\rm dB $, $ {p_B} = {p_E} = 0.5 $, $ {m_B} = 5$, ${m_E} = 0.5 $ and  ${K_{B,1}}{\rm{/}}{K_{B,2}} = {K_{E,1}}{\rm{/}}{K_{E,2}} = 5$. It can be easily observed that SOP decreases as ${\overline K _B}$ increases. Figure \ref{pnz_ki} plots the PNZ versus ${\overline \gamma  _B}$, for  $ {\overline \gamma  _E} = 4$  $\rm  dB$, ${p_B} = {p_E} = 0.5$, ${m_B} = {m_E} = 0.5$, ${K_{B,1}} = 100$, ${K_{B,2}} = 10$,  $ {K_{E,1}}{\rm{/}}{K_{E,2}}=10 $. It can be observed that PNZ increases as ${\overline K _E}$ increases.
		This is due to the reason that the large power of LoS components increases the average SNR.
		Again, perfect agreement is observed between analytical results and Monte-Carlo simulations. Furthermore, the asymptotic results of SOP and PNZ gradually approach the analytical results as $ {\bar \gamma_B } $ increases.
		
		\section{Conclusion}
 We investigate the PLS over the ARS fading channel. More specifically, we propose the expressions of SOP, PNZ and ASC for two cases: $m$ is a positive real number and $m$ is a positive integer number. In addition, we derive the asymptotic expressions of ASC, SOP and PNZ which all match well the exact ones at high-SNR values, respectively.
 To this end, numerical results have been presented to validate the proposed analytical expressions. The important insights we provided are useful. For example, the performance of the considered system can be improved by increasing the average SNR of the main channel or decreasing the average SNR of the eavesdropper channel. Besides, SOP  decreases as ${\overline K _B}$ or ${m_B}$ increases.

		\begin{appendices}
		\renewcommand{\theequation}{A-\arabic{equation}}
	\section{Proof of theorem \ref{prop1}}\label{AppendixA}
		\setcounter{equation}{0}
		Substituting (\ref{11}) and (\ref{12}) into (\ref{18}), we have
		{\small \begin{align}
		 &{I_1}= \int_0^\infty  {\ln \left( {1 + {\gamma _B}} \right){f_B}\left( {{\gamma _B}} \right){F_E}\left( {{\gamma _B}} \right)d{\gamma _B}}\notag\\
		&  = \int_0^\infty  {\ln \left( {1 + {\gamma _B}} \right)\left( {{p_B}{f_{RS,B,1}}\left( {{\gamma _B}} \right) + \left( {1 - {p_B}} \right){f_{RS,B,2}}\left( {{\gamma _B}} \right)} \right)}\notag\\
		&\quad\times
		\left( {{p_E}{F_{RS,E,1}}\left( {{\gamma _B}} \right) + \left( {1 - {p_E}} \right){F_{RS,E,2}}\left( {{\gamma _B}} \right)} \right)d{\gamma _B}\notag\\
		& = {R_{1,1}} + {R_{1,2}} + {R_{2,1}} + {R_{2,2}},
		\end{align}}
		where
		{\small \begin{equation}\label{A-2}
		{R_{i,j}}\!\! =\! {Q_{\!B\!,i}}{Q_{\!E\!,j}}\!\!\!\int_0^\infty \!\!\!\! {\ln } \left( {1\! + \!{\gamma _B}} \right)\!{f_{\!RS,B,i}}\!\left( {{\gamma _B}} \right)\!{F_{\!RS,E,j}}\!\left( {{\gamma _B}} \right)\!d{\gamma _B}.
		\end{equation}
		}
        In order to solve \eqref{A-2}, we re-express ${\rm ln}(1+x) $ in terms of the Meijer's $ G $-function as \cite[Eq. (8.4.6.5)]{prudnikov1986integrals}
		 {\small \begin{equation}\label{A-6}
\ln \left( {1 + x} \right) = G_{2,2}^{1,2}\left( {x\left| {\begin{array}{*{20}{c}}
		{1,1}\\
		{1,0}
		\end{array}} \right.} \right)= G_{2,2}^{2,1}\left( {\frac{1}{x}\left| {\begin{array}{*{20}{c}}
		{0,1}\\
		{0,0}
	\end{array}} \right.} \right).
		\end{equation}}
		With the help of \cite[Eq. (35) and Eq. (39)]{al2019performance}, the confluent Lauricella hypergeometric function  $ {\Phi _2}\left(  \cdot  \right) $ can be written as
		{\small \begin{align}\label{A-7}
		&{\Phi _2}\left( {{a_1},{a_2};b; - {x_1}t, - {x_2}t} \right) = \frac{{\Gamma \left( b \right)}}{{\Gamma \left( {{a_1}} \right)\Gamma \left( {{a_2}} \right)}}\frac{1}{{{{\left( {2\pi i} \right)}^2}}}\notag\\
		&\times\!\!\!\int_{{\!\Re_{\!1}}} \!\!\!{\int_{{\!\Re_{\!2}}}\!\!\!\! {\frac{{\Gamma \!\left( {{r_1}} \right)\!\Gamma \!\left( {{a_1}\!\! -\! \!{r_1}} \right)\!\Gamma \!\left( {{r_2}} \right)\!\Gamma \!\left( {{a_2}\! \!-\!\! {r_2}} \right)}}{{\Gamma \!\left( {b\! -\! {r_1} \!-\! {r_2}} \right)}}\!{{\left( \!{{x_{\!1}}{t_{\!1}}} \!\right)}^{ \!- \!{r_{\!1}}}}\!{{\left( \!{{x_{\!2}}{t_{\!2}}} \!\right)}^{\! -\! {r_{\!2}}}}\!d{r_{\!1}}\!d{r_{\!2}}} } .
		\end{align}	}		
		Using \cite[Eq. (9.210.1) and Eq. (9.212.1)]{gradshteyn2007} and \cite[07.20.07.0003.01]{web}, we can rewrite $ {}_1{F_1}\left( . \right) $ as
	 {\small \begin{equation}\label{A-8}
		{}_1{F_1}\left( {\alpha ,\!\gamma ;z} \right)\! =\! \frac{{\exp (z)\Gamma \left( \gamma  \right)}}{{2\pi i\Gamma \left( {\gamma \! - \!\alpha } \right)}}\int_{_\mathcal{L}} {\frac{{\Gamma \left( s \right)\Gamma \left( {\gamma  \!-\! \alpha \! -\! s} \right)}}{{\Gamma \left( {\gamma  \!-\! s} \right)}}{{\left( z \right)}^{ - s}}ds} .
		\end{equation}}
		Substituting (\ref{9}) and (\ref{10}) into (\ref{A-2}) and with the help of (\ref{A-6}), (\ref{A-7}) and (\ref{A-8}) , we can express $ {R_{1,1}} $  as \eqref{A-9} at the top of the next page.	
		\newcounter{mytemp6}
		\begin{figure*}[t]
			\normalsize
			\setcounter{mytemp6}{\value{equation}}
			\setcounter{equation}{5}
			{\small \begin{align}\label{A-9}
				&{R_{1,1}}{\rm{ = }}{p_B}{p_E}\frac{{1 + {{\overline K }_E}}}{{{{\overline \gamma  }_E}}}{\left( {\frac{{{m_E}}}{{{m_E} + {K_{1,E}}}}} \right)^{{m_E}}}\frac{{1 + {{\overline K }_B}}}{{{{\overline \gamma  }_B}}}{\left( {\frac{{{m_B}}}{{{m_B} + {K_{1,B}}}}} \right)^{{m_B}}}\frac{1}{{\Gamma \left( {1 - {m_B}} \right)\Gamma \left( {1 - {m_E}} \right)\Gamma \left( {{m_E}} \right)}}\notag\\
				&\times\!\!\frac{1}{{{{\left( {2\pi i} \right)}^4}}}\int_{{\Re_1}} {\int_{{\Re_2}} {\int_{{\Re_3}} {\int_{{\Re_4}} {\frac{{\Gamma \left( {{r_1}} \right)\Gamma \left( {1 - {m_E} - {r_1}} \right)\Gamma \left( {{r_2}} \right)\Gamma \left( {{m_E} - {r_2}} \right)\Gamma \left( {{r_3}} \right)\Gamma \left( {1 - {m_B} - {r_3}} \right)\Gamma \left( {1 - {r_4}} \right)\Gamma \left( {{r_4}} \right)\Gamma \left( {{r_4}} \right)}}{{\Gamma \left( {2 - {r_1} - {r_2}} \right)\Gamma \left( {1 - {r_3}} \right)\Gamma \left( {1{\rm{ + }}{r_4}} \right)}}} } } }\notag\\
				&\times\!\! {\left(\! {\frac{{1\! +\! {{\overline K }_E}}}{{{{\overline \gamma  }_E}}}} \!\right)^{\!\!\! - {r_1}}}\!\!{\left(\! {\frac{{{m_E}\!\left( {1\! + \!{{\overline K }_E}} \right)}}{{{{\overline \gamma  }_E}\left( {{m_E}\! +\! {K_{1,E}}} \right)}}}\! \right)^{ \!\!\!- {r_2}}}\!\!{\left(\! {\frac{{{K_{1,B}}\!\left( {1\! +\! {{\overline K }_B}} \right)}}{{{{\overline \gamma  }_B}\!\left(\! {{m_B} \!+\! {K_{1,B}}} \!\right)}}} \right)^{\!\!\! - {r_3}}}\!\!\!\!\underbrace {\int_0^\infty  \!\!\!\!{{\gamma _B}^{1 \!- {r_1}\! - {r_2}\! - {r_3} \!+ {r_4}}\!\exp \!\left( \!{ - \frac{{{m_B}\left( {1 + {{\overline K }_B}} \right)}}{{{{\overline \gamma  }_B}\left( {{m_B} + {K_{1,B}}} \right)}}{\gamma _B}}\! \right)\!d{\gamma _B}} }_{{\kappa _1}}d{r_4}d{r_3}d{r_2}d{r_1}.
				\end{align}}
			\hrulefill	
			{\small \begin{align}\label{A-11}
				&	{R_{1,1}}{\rm{ = }}{p_B}{p_E}\frac{{{{\overline \gamma  }_B}\left( {1 + {{\overline K }_E}} \right)}}{{{{\overline \gamma  }_E}\left( {1 + {{\overline K }_B}} \right)}}{\left( {\frac{{{m_E}}}{{{m_E} + {K_{1,E}}}}} \right)^{{m_E}}}{\left( {\frac{{{m_B}}}{{{m_B} + {K_{1,B}}}}} \right)^{{m_B} - 2}}\frac{1}{{\Gamma \left( {1 - {m_B}} \right)\Gamma \left( {1 - {m_E}} \right)\Gamma \left( {{m_E}} \right)}}\notag\\
				&\times H_{1,3:1,1;1,1;1,1;2,1}^{0,1:1,1;1,1;1,1;1,2}\left[ {\left. {\frac{{{{\overline \gamma  }_B}\left( {{m_B} + {K_{1,B}}} \right)\left( {1 + {{\overline K }_E}} \right)}}{{{{\overline \gamma  }_E}{m_B}\left( {1 + {{\overline K }_B}} \right)}},\frac{{{{\overline \gamma  }_B}{m_E}\left( {{m_B} + {K_{1,B}}} \right)\left( {1 + {{\overline K }_E}} \right)}}{{{{\overline \gamma  }_E}{m_B}\left( {{m_E} + {K_{1,E}}} \right)\left( {1 + {{\overline K }_B}} \right)}},\frac{{{K_{1,B}}}}{{{m_B}}},\frac{{{{\overline \gamma  }_B}\left( {{m_B} + {K_{1,B}}} \right)}}{{{m_B}\left( {1 + {{\overline K }_B}} \right)}}} \right|} \right.\notag\\
				&\left. {\begin{array}{*{20}{c}}
					{\left( { - 1;1,1,1,1} \right)}\\
					{\left( { - 1;1,1,0,0} \right),\left( {0;0,0,1,0} \right),\left( {0;0,0,0,1} \right)}
					\end{array}\begin{array}{*{20}{c}}
					:\\
					:
					\end{array}\begin{array}{*{20}{c}}
					{\left( {{m_E},1} \right)}\\
					{\left( {0,1} \right)}
					\end{array}\begin{array}{*{20}{c}}
					;\\
					;
					\end{array}\begin{array}{*{20}{c}}
					{\left( {1 - {m_E},1} \right)}\\
					{\left( {0,1} \right)}
					\end{array}\begin{array}{*{20}{c}}
					;\\
					;
					\end{array}\begin{array}{*{20}{c}}
					{\left( {{m_B},1} \right)}\\
					{\left( {0,1} \right)}
					\end{array}\begin{array}{*{20}{c}}
					;\\
					;
					\end{array}\begin{array}{*{20}{c}}
					{\left( {1,1} \right),\left( {1,1} \right)}\\
					{\left( {1,1} \right)}
					\end{array}} \right].
				\tag{A-8}
				\end{align}}
			\hrulefill
			{\small \begin{align}\label{A-12}
				&{I_3} \!=\! \frac{{{p_{\!E}}\left( {1\!\! +\!\! {{\overline K }_{\!E}}} \right)}}{{{{\overline \gamma  }_{\!E}}\Gamma\! \left( {1\!\! -\!\! {m_{\!E}}} \right)}}\!{\left(\! {\frac{{{m_E}}}{{{m_{\!E}}\! +\! {K_{\!E,1}}}}}\! \right)^{\!\!{m_{\!E}}}}\!\!{\left(\!\! {\frac{1}{{2\pi i}}} \!\!\right)^{\!2}}\!\!\!\!\!\int_{\!{\Re_{\!1}}} \!\!{\int_{\!{\Re_{\!2}}} \!\!\!\!\!{\frac{{\Gamma\! \left( {{r_{\!1}}} \right)\!\Gamma \!\left(\! {1 \!\!-\!\! {m_{\!E}}\!\! -\! \!{r_{\rm{1}}}}\! \right)\!\Gamma\! \left(\! {1 \!- \!{r_2}}\! \right)\!{\Gamma^2}\! \left( {{r_{\!2}}} \right)\!}}{{\Gamma \left( {1 - {r_1}} \right)\Gamma \left( {1{\rm{ + }}{r_2}} \right)}}} } \!{\left( \!\!{\frac{{{K_{\!1,E}}\!\left( {1 \!+\! {{\overline K }_E}} \right)}}{{{{\overline \gamma  }_{\!E}}\!\left(\!\! {{m_{\!E}} \!\!+\!\! {K_{\!1,E}}}\!\! \right)}}} \!\!\right)^{\!\! \!\!- \!{r_1}}}
				\!\!\!\!\!\!\underbrace {\int_0^\infty  \!\!\!\!\!\!{\exp \left( \!\!{ - \frac{{{m_{\!E}}\left( \!{1\!\! +\!\! {{\overline K }_{\!E}}} \!\right)}}{{{{\overline \gamma  }_{\!E}}\!\left(\! {{m_{\!E}}\!\! +\!\! {K_{\!E,1}}}\! \right)}}\!{\gamma _{\!E}}}\!\! \right)\!{\gamma _{\!E}}^{{r_{\!2}}\! -\! {r_{\!1}}}\!d{\gamma _{\!E}}} }_{{\kappa _2}}\!d{r_2}d{r_1}\notag\\
				&+\!\! \frac{{\left( {1\!\! -\!\! {p_{\!E}}} \right)\!\!\left( {1\!\! +\!\! {{\overline K }_{\!E}}} \right)}}{{{{\overline \gamma  }_{\!E}}\Gamma \left( {1 \!\!-\!\! {m_{\!E}}} \right)}}\!{\left( \!{\frac{{{m_E}}}{{{m_{\!E}}\! \!+\!\! {K_{\!2,E}}}}}\! \right)^{\!\!\!{m_{\!E}}}}\!\!\!{\left( \!\!{\frac{1}{{2\pi i}}}\! \!\right)^{\!\!\!2}}\!\!\!\!\int_{{\Re_{\!1}}} \!\!{\int_{{\Re_{\!2}}}\!\!\!\!\! {\frac{{\Gamma\! \left( {{r_{\!1}}} \right)\!\Gamma \!\left( \!{1\!\! -\! \!{m_{\!E}} \!\!-\! \!{r_{\rm{1}}}} \right)\!\Gamma \!\left( {1\! \!-\!\! {r_{\!2}}} \right)\!{\Gamma^2}\! \left( {{r_{\!2}}} \right)}}{{\Gamma \left( {1 - {r_1}} \right)\Gamma \left( {1{\rm{ + }}{r_2}} \right)}}} }\!\! {\left(\!\! {\frac{{{K_{\!2,E}}\!\left( {1 \!\!+\!\! {{\overline K }_{\!E}}} \right)}}{{{{\overline \gamma  }_{\!E}}\!\left( {{m_{\!E}}\!\! +\!\! {K_{\!2,E}}} \right)}}} \!\!\right)^{ \!\!\!\!- \!{r_1}}}
				\!\!\!\!\!\!\underbrace {\int_0^\infty  \!\!\!\!\!\!{\exp\! \!\left(\! \!{ - \frac{{{m_{\!E}}\!\left(\! {1\!\! +\!\! {{\overline K }_{\!E}}}\! \right)}}{{{{\overline \gamma  }_{\!E}}\!\left( {{m_{\!E}} \!\!+\!\! {K_{\!2,E}}} \right)}}\!{\gamma _{\!E}}} \!\!\right){\gamma _{\!E}^{{r_2}\! -\! {r_1}}}\!d{\gamma _{\!E}}} }_{{\kappa _3}}d{r_2}d{r_1}\!.
				\tag{A-9}
				\end{align}	}
			\hrulefill
			\setcounter{equation}{\value{mytemp6}}
		\end{figure*}
		\setcounter{equation}{6}
		With the help of \cite[Eq. (3.381.4)]{gradshteyn2007},  $ {\kappa _1} $  can be expressed in closed-form as
	{\small 	\begin{equation}\label{A-10}
		{\kappa _1} \!\!=\!\! {\left( \!{\frac{{{m_{\!B}}\!\left( {1\! +\! {{\overline K }_{\!B}}} \right)}}{{{{\overline \gamma  }_{\!B}}\!\left( \!{{m_{\!B}}\! +\! {K_{\!1,B}}}\! \right)}}} \!\right)^{\!{r_1} + {r_2}\! + {r_3} \!- {r_4} \!- 2}}\!\!\Gamma \!\left( {2 \!- \!{r_1} \!- \!{r_2}\! -\! {r_3}\! +\! {r_4}} \right).
		\end{equation}}
		Substituting \eqref{A-10} into \eqref{A-9} and with the help of the definition of the multivariate $ H $-function \cite[Eq. (28)]{alhennawi2015closed}, we derive 	\eqref{A-11} at the top of	the next page.
			

		\setcounter{equation}{9}
		After the same mathematical derivation process as $ {R_{1,1}} $, it is easy to derive similar closed-form expression of $ {R_{1,2}} $, $ {R_{2,1}} $ and $ {R_{2,2}} $. Then, we can obtain the exact expression of $ {R_{i,j}} $  as (\ref{24}).
		
	Following similar steps, we  derive (\ref{19}) by replacing each $ B $ and $E$ with $E$ and $B$, and the exact expression of $ {T_{i,j}} $ is expressed as \eqref{25}.
		
		For $ {I_3} $ , we substitute (\ref{11}) into (\ref{20}) and using \eqref{A-6} and \eqref{A-8} to yield \eqref{A-12}at the top of	the next page.
		With the help of \cite[Eq. (3.381.4)]{gradshteyn2007}, $ {\kappa _2} $  and $ {\kappa _3} $  can be  expressed as
		{\small  \begin{equation}\label{A-13}
		{\kappa _2} = {\left( {\frac{{{m_E}\left( {1 + {{\overline K }_E}} \right)}}{{{{\overline \gamma  }_E}\left( {{m_E} + {K_{1,E}}} \right)}}} \right)^{{r_1} - {r_2} - 1}}\Gamma \left( {1 - {r_1} + {r_2}} \right),
		\end{equation}
		\begin{equation}\label{A-14}
		{\kappa _3} = {\left( {\frac{{{m_E}\left( {1 + {{\overline K }_E}} \right)}}{{{{\overline \gamma  }_E}\left( {{m_E} + {K_{2,E}}} \right)}}} \right)^{{r_1} - {r_2} - 1}}\Gamma \left( {1 - {r_1} + {r_2}} \right).
		\end{equation}}	
		Substituting \eqref{A-13} and \eqref{A-14} into (\ref{A-12}) and with the help of the definition of the multivariate Fox's  $ H $-function \cite[Eq. (28)]{alhennawi2015closed}, we can derive the closed-form expression of $ {I_3} $ as (\ref{23}).
		
		\section{Proof of Corollary \ref{prop2}}\label{AppendixB}
		\renewcommand{\theequation}{B-\arabic{equation}}
		\setcounter{equation}{0}
		The  expression of the ASR fading model's PDF is given by \cite[Eq. (13)]{fernandez2019tractable}, and the CDF can be obtained with the help of \cite[Eq. (17)]{fernandez2019tractable} as
 		{\small \begin{align}\label{B-1}
		&{f_\ell }\left( \gamma  \right)\! = p\sum\limits_{n = 0}^{{m_\ell } - 1} {{B_{n,\ell ,1}}\frac{{{\gamma ^{{m_\ell } - n - 1}}}}{{\overline \rho  _{\ell ,1}^{{m_\ell } - n}\Gamma \left( {{m_\ell } - n} \right)}}\exp \left( { - \frac{\gamma }{{{{\overline \rho  }_{\ell ,1}}}}} \right)} \notag\\
		& + \left( {1 - p} \right)\sum\limits_{n = 0}^{{m_\ell } - 1} {{B_{n,\ell ,2}}\frac{{{\gamma ^{{m_\ell } - n - 1}}}}{{\overline \rho  _{\ell ,2}^{{m_\ell } - n}\Gamma \left( {{m_\ell } - n} \right)}}\exp \left( { - \frac{\gamma }{{{{\overline \rho  }_{\ell ,2}}}}} \right)}  ,
		\end{align}
 		\begin{align}\label{B-2}
		&{F_\ell }\left( \gamma  \right)\!\!= p\sum\limits_{j = 0}^{{m_\ell } - 1} \!\!{{B_{j,\ell ,1}}\left( {1 - \exp \left( { - \frac{\gamma }{{{{\bar \rho }_{\ell ,1}}}}} \right)\sum\limits_{k = 0}^{{m_\ell } - j - 1} {\frac{{{\gamma ^k}}}{{\bar \rho _{\ell ,1}^kk!}}} } \right)}\!\! \notag\\
		&+\! \left( {1 \!-\! p} \right)\!\!\sum\limits_{j = 0}^{{m_\ell } - 1} \!\!{{B_{j,\ell ,2}}\left( {1 - \exp \left( { - \frac{\gamma }{{{{\bar \rho }_{\ell ,2}}}}} \right)\sum\limits_{k = 0}^{{m_\ell } - j - 1} {\frac{{{\gamma ^k}}}{{\bar \rho _{\ell ,2}^kk!}}} } \right)} ,
		\end{align}}
		where $ {\rho _{\ell ,r}} $ and $ {B_{n,\ell ,r}} $ are   defined as \eqref{30} and \eqref{31} respectively. By comparing \eqref{B-1} with \eqref{7}, \eqref{B-2} with \eqref{8}, the PDF and CDF of Rician shadowed model can be expressed as
	{\small  \begin{align}\label{B-3}
		{f_{RS,\ell ,r}}\left( \gamma  \right) \!=\!\! \sum\limits_{n = 0}^{{m_\ell } - 1}\!\! {{B_{n,\ell ,r}}\frac{{{\gamma ^{{m_\ell } - n - 1}}}}{{\overline \rho  _{\ell ,r}^{{m_\ell } - n}\Gamma \left( {{m_\ell } - n} \right)}}\exp \left( { - \frac{\gamma }{{{{\overline \rho  }_{\ell ,r}}}}} \right)} ,
		\end{align}
		\begin{align}\label{B-4}
		{F_{RS,\ell ,r}}\left( \gamma  \right) \!=\!\! \sum\limits_{j = 0}^{{m_\ell } - 1} \!\!{{B_{j,\ell ,r}}\!\left(\! {1 \!- \!\exp\! \left( { \!-\! \frac{\gamma }{{{{\bar \rho }_{\ell ,r}}}}} \!\right)\!\!\sum\limits_{k = 0}^{{m_\ell } - j - 1} \!\!{\frac{{{\gamma ^k}}}{{\bar \rho _{\ell ,r}^kk!}}} } \right)} .
		\end{align}	}
		
		For integer values of $ m $, substituting \eqref{B-3} and \eqref{B-4} into \eqref{A-2}, we can express $ R_{1,1} $ as
		{\small  \begin{align}\label{B-5}
&{R_{1,1}}\! =\! {p_B}{p_E}\sum\limits_{l = 0}^{{m_E} - 1} {\sum\limits_{n = 0}^{{m_B} - 1} \left( {{{B_{l,E,1}}{B_{n,B,1}}\frac{1}{{\overline \rho  _{B,1}^{{m_B} - n}\Gamma \left( {{m_B} - n} \right)}}}} \right. }\notag\\
&\quad\times\int_0^\infty  {\ln \left( {1 + {\gamma _B}} \right)}\gamma _B^{{m_B} - n - 1}\left. {\exp \left( { - \frac{{{\gamma _B}}}{{{{\overline \rho  }_{1,B}}}}} \right)d{\gamma _B}} \right)\notag\\
& \quad- {p_B}{p_E}\sum\limits_{l = 0}^{{m_E} - 1} {\sum\limits_{n = 0}^{{m_B} - 1} {\sum\limits_{k = 0}^{{m_E} - l - 1}\!\!\!
		\left( {{\frac{{{B_{l,E,1}}{B_{n,B,1}}}}{{\overline \rho  _{B,1}^{{m_B} - n}\overline \rho  _{E,1}^k\Gamma \left( {{m_B} - n} \right)k!}}} } \right.} }\notag\\
&\quad\times\!\left. {\!\!\!\int_0^\infty  \!\!{\ln\! \left( {1 \!+ \!{\gamma _B}} \right)\!\gamma _B^{k\! +\! {m_B}\! -\! n\! - \!1}\!\exp \!\left( \!{ - \frac{{{{\overline \rho  }_{B,1}}\! + \!{{\overline \rho  }_{E,1}}}}{{{{\overline \rho  }_{B,1}}{{\overline \rho  }_{E,1}}}}{\gamma _B}}\! \right)\!d{\gamma _B}}}\!\! \right) \!.
		\end{align}	}
		With the help of \eqref{A-6} and \cite[Eq. (7.813.1)]{gradshteyn2007}, $ R_{1,1} $ can be expressed as
		{\small \begin{align}\label{B_R11}
			{R_{1,1}}\! &=\! {p_{\!B}}{p_{\!E}}\!\!\!\sum\limits_{l = 0}^{{m_{\!E}} \!-\! 1} \!{\sum\limits_{n = 0}^{{m_B} - 1} \!\!{\frac{{{B_{l,E,1}}{B_{n,B,1}}}}{{\Gamma \left( {{m_B} - n} \right)}}G_{3,2}^{1,3}} } \left( {{{\overline \rho  }_{B,1}}\left|\!\!\!\! {\begin{array}{*{20}{c}}
					{1\! -\! {m_B} \!+\! n,1,1}\\
					{1,0}
					\end{array}} \right.}\!\!\!\!\! \right)\notag\\
			&-\! {p_{\!B}}{p_{\!E}}\!\!\!\sum\limits_{l = 0}^{{m_{\!E}}\! -\! 1} {\sum\limits_{n = 0}^{{m_{\!B}} \!-\! 1} \!{\sum\limits_{k = 0}^{{m_{\!E}} \!-\! l\! -\! 1}\!\!\! {\frac{{{B_{l,E,1}}{B_{n,B,1}}\overline \rho  _{B,1}^k\overline \rho  _{E,1}^{{m_B} - n}}}{{\Gamma \!\left( {{m_{\!B}} \!-\! n} \right)k!{{\left( {{{\overline \rho  }_{\!E,1}}\! +\! {{\overline \rho  }_{\!B,1}}} \right)}^{k + {m_{\!B}} - n}}}}} } }\notag\\
			&\quad\times G_{3,2}^{1,3}\left(\! {\frac{{{{\overline \rho  }_{B,1}}{{\overline \rho  }_{E,1}}}}{{{{\overline \rho  }_{B,1}} + {{\overline \rho  }_{E,1}}}}\left|\!\!\! {\begin{array}{*{20}{c}}
					{ - k - {m_B} + n + 1,1,1}\\
					{1,0}
					\end{array}} \right.} \!\!\!\right) .
			\end{align} }
%
Following the same  derivation process as $ {R_{1,1}} $, we can easily derive the  similar closed-form expressions of $ {R_{1,2}} $, $ {R_{2,1}} $ and $ {R_{2,2}} $. Then we obtain $ {R_{i,j}} $  as \eqref{28}.
		Besides, $ {I_2} $ can be deduced  by replacing each $ B $ and $E$ in $ I_1 $ with $E$ and $B$. The  general expression of $ {T_{i,j}} $ can be obtained as \eqref{29}.
		
		For $ {I_3} $ , we substitute \eqref{B-1} into \eqref{20} to yield
	 {\small   \begin{align}\label{B-8}
		 &{I_3}\! ={p_{\!E}}\!\!\!\sum\limits_{n = 0}^{{m_{\!E}}\! -\! 1}\!\!\! \left( {{\frac{{{B_{n,E,1}}}}{{\overline \rho  _{\!E,1}^{{m_{\!E}} \!-\! n}\Gamma \!\left(\! {{m_{\!E}}\! -\! n}\! \right)}}}} \right.\!\!\!\left. {\int_0^\infty \!\!\! \!\!{\ln\! \left( {1\!\! + \!\!{\gamma _{\!E}}} \right)\!\gamma _E^{{m_{\!E}}\! -\! n\! -\! 1}\exp\! \left(\! \!{ - \frac{{{\gamma _{\!E}}}}{{{{\overline \rho  }_{\!E,1}}}}}\!\! \right)\!d{\gamma _{\!E}}}}\! \right) \notag\\
		&+\!\left(\! {1 \!\!-\!\! {p_ {\!E}}}\! \right)\!\!\!\sum\limits_{n = 0}^{{m_{\!E}} \!-\! 1}\!\!\! \left(\! {{\frac{{{B_{n,E,2}}}}{{\overline \rho  _{E,2}^{{m_{\!E}}\! -\! n}\Gamma\! \left(\! {{m_{\!E}}\! -\! n} \!\right)}} }} \right.\!\!\!\!\left. {\int_0^\infty  {\!\!\!\!\!\!\ln\! \left(\! {1\! +\! {\gamma _{\!E}}} \!\right)\!\gamma _{\!E}^{{m_{\!E}}\! -\! n\! -\! 1}\!\exp\!\! \left(\!\! { - \frac{{{\gamma _E}}}{{{{\overline \rho  }_{\!E,2}}}}}\!\! \right)\!\!d{\gamma _{\!E}}}} \!\!\right)\!.
		\end{align}}
		With the help of \eqref{A-6} and \cite[Eq. (7.813.1)]{gradshteyn2007}, the two integrals in \eqref{B-8} can be solved, and we obtain the  close-form expressions of $ {I_3} $ as \eqref{27}.
		
		\section{Proof of proposition \ref{prop111}}\label{AppendixCNEW}
				\renewcommand{\theequation}{C-\arabic{equation}}
				\setcounter{equation}{0}
With the help of \cite[Eq. (28)]{alhennawi2015closed},
the multivariate Fox's $H$-function $ {H_R} $ in \eqref{24} can be
expressed in terms of the Mellin-Barnes integral as \eqref{CNEW1} at the bottom of the next page.
\newcounter{mytemp8}
\begin{figure*}[b]
	\normalsize
	\setcounter{mytemp8}{\value{equation}}
	\hrulefill
	\setcounter{equation}{0}
	{\small  \begin{align} \label{CNEW1}
		&{H_R} =\frac{1}{{{{\left( {2\pi i} \right)}^{\rm{4}}}}}\int_{{\Re _1}} {\int_{{\Re _2}} {\int_{{\Re _3}} {\int_{{\Re _4}} { {\frac{{\Gamma \left( {2 - {r_1} - {r_2} - {r_3} + {r_4}} \right)}}{{\Gamma \left( {2 - {r_1} - {r_2}} \right)\Gamma \left( {1 - {r_3}} \right)\Gamma \left( {1{\rm{ + }}{r_4}} \right)}}\Gamma \left( {{r_1}} \right)\Gamma \left( {1 \!-\! {m_E}\! -\! {r_1}} \right)\Gamma \left( {{r_2}} \right)\Gamma \left( {{m_E}\! -\! {r_2}} \right)} } } } }\Gamma \left( {{r_3}} \right)\Gamma \left( {1\! -\! {m_B}\! - \!{r_3}} \right)\notag\\
		&\times\!\Gamma \!\left( \!{1\! \!-\!\! {r_{\!4}}}\! \right)\!\Gamma \!\left( {{r_{\!4}}} \right)\!\Gamma\! \left( {{r_{\!4}}} \right)\!{\left(\! {\frac{{{{\overline \gamma  }_E}{m_B}\left( {1 + {{\overline K }_B}} \right)}}{{{{\overline \gamma  }_{\!B}}\!\left(\! {{m_{\!B}}\!\! +\!\! {K_{\!B,i}}} \!\right)\!\left( \!{1\!\! +\!\! {{\overline K }_{\!E}}}\! \right)}}} \!\right)^{\!\!\!{r_1}}}\!\!\! {{\left( {\frac{{{{\overline \gamma  }_E}{m_B}\left( {{m_E} + {K_{E,j}}} \right)\left( {1 + {{\overline K }_B}} \right)}}{{{{\overline \gamma  }_B}{m_E}\left( {{m_B} + {K_{B,i}}} \right)\left( {1 + {{\overline K }_E}} \right)}}} \right)^{\!\!\!{r_2}}}\!\!\!{\left(\! {\frac{{{m_B}}}{{{K_{B,i}}}}} \!\right)^{  \!\!\!{r_3}}}\!\!{\left(\! {\frac{{{{\overline \gamma  }_B}\left( {{m_B} + {K_{B,i}}} \right)}}{{{m_B}\left( {1 + {{\overline K }_B}} \right)}}}\! \right)^{\!\!{r_4}}}}\!\!\! d{r_4}d{r_3}d{r_2}d{r_1}.
		\end{align}}
\hrulefill		
		{\small 		\begin{align}\label{ss2}
			&{W_{i,j}}\! \!=\! {Q_{B\!,i}}{Q_{E\!,j}}\frac{{\left( {1\! +\! {{\overline K }_B}} \right)\!\!\left( {1 \!+\! {{\overline K }_E}} \right)}}{{{{\overline \gamma  }_B}{{\overline \gamma  }_E}\Gamma \left( {1\! - \!{m_E}} \right)}}\!\!{\left(\! {\frac{{{m_B}}}{{{m_B} \!+\! {K_{B,i}}}}} \!\right)^{\!\!{m_B}}}\!\!\!{\left(\! {\frac{{{m_E}}}{{{m_E}\! +\! {K_{E,j}}}}} \!\right)^{{\!\!m_E}}}\! \!\!\!\!\!\!\!\sum\limits_{{n_1} = {n_2} = 0}^\infty\!\!\!  {\left( {\frac{{{{\left( {1\! - \!{m_B}} \right)}_{{n_1}}}\!{{\left( {{m_B}} \right)}_{{n_2}}}}}{{{n_1}!{n_2}!{{\left( 2 \right)}_{{n_1} + {n_2}}}}}{{\left(\!\! { - \frac{{1 \!+\! {{\overline K }_B}}}{{{{\overline \gamma  }_B}}}} \!\right)}^{{n_1}}}\!\!\!{{\left(\! \!{ - \frac{{{m_B}\left( {1 + {{\overline K }_B}} \right)}}{{{{\overline \gamma  }_B}\left( {{m_B} \!+ \!{K_{B,i}}} \!\right)}}} \right)}^{{\!n_2}}}} \right.} \notag\\ &\times\frac{1}{{2\pi i}}\int_\Re  {\frac{{\Gamma \left( r \right)\Gamma \left( {1 - {m_E} - r} \right)}}{{\Gamma \left( {1 - r} \right)}}\left( {\frac{{{K_{E,j}}\left( {1 + {{\overline K }_E}} \right)}}{{{{\overline \gamma  }_E}\left( {{m_E} + {K_{E,j}}} \right)}}} \right)^{ - r}}\!\!\! \left. { \underbrace {\int_{\rm{0}}^\infty \!\!\! {{{\left( {{R_s}{\gamma _E} + w} \right)}^{1 + {n_1} + {n_2}}}\gamma _E^{ - r}\exp \left( { - \frac{{\left( {1 + {{\overline K }_E}} \right){m_E}}}{{{{\overline \gamma  }_E}\left( {{m_E} + {K_{E,j}}} \right)}}{\gamma _E}} \right)d{\gamma _E}} }_{{\kappa _{\rm{4}}}}dr} \right).
						\tag{D-4}
			\end{align}}
		\hrulefill
		{\small \begin{align}\label{C-3}
			{W_{i,j}}\!\! = &{Q_{B,i}}{Q_{E,j}}\!\!\!\sum\limits_{l = 0}^{{m_{\!B}} \!- 1} \!{\sum\limits_{n = 0}^{{m_{\!E}}\! - 1}\!\! {\frac{{{B_{l,B,i}}{B_{n,E,j}}}}{{\bar \rho _{E,j}^{{m_E}\! -\! n}\Gamma ({m_E} - n)}}\!\!\underbrace {\int_0^\infty\! \!\!\!\!\! {{\gamma _E}^{{m_E} \!- n \!- 1}\!\exp\! \left(\! { - \frac{{{\gamma _E}}}{{{{\bar \rho }_{E,j}}}}} \!\right)\!d{\gamma _E}} }_{{\kappa _5}}} }\! -\! {Q_{B,i}}{Q_{E,j}}\!\!\!\sum\limits_{l = 0}^{{m_B}\!- \!1}\! {\sum\limits_{n = 0}^{{m_E}\! - \!1} \!{\sum\limits_{k = 0}^{{m_B} \!- l\! - 1}\!\! {\sum\limits_{t = 0}^k\Bigg(\!\! \exp\!\! \left( { - \frac{{{R_s} - 1}}{{{{\overline \rho  }_{B,i}}}}} \right) } } }\notag\\
			&\quad \times	{\frac{{{B_{l,B,i}}{B_{n,E,j}}{{\left( {{R_s} - 1} \right)}^t}R_s^{k - t}}}{{\overline \rho  _{B,i}^k\overline \rho  _{E,j}^{{m_E} - n}\Gamma ({m_E} - n)t!\left( {k - t} \right)!}}}
			\underbrace {\int_0^\infty  {{\gamma _E}^{{m_E} - 1 - n + k - t}\exp \left( { - \frac{{{{\overline \rho  }_{B,i}} + {R_s}{{\overline \rho  }_{E,j}}}}{{{{\overline \rho  }_{E,j}}{{\overline \rho  }_{B,i}}}}{\gamma _E}} \right)d{\gamma _E}} }_{{\kappa _6}}\Bigg),
			\tag{D-7}
			\end{align} }
	\setcounter{equation}{\value{mytemp8}}
\end{figure*}
\setcounter{equation}{1}
  Equation \eqref{CNEW1} can
be further derived by calculating the residue of the related
integrand at the nearest pole to the integration contour \cite{chergui2016performance}. When $ {\overline \gamma  _B} \to \infty   $,
we have $\frac{{{{\overline \gamma  }_E}{m_B}\left( {1 + {{\overline K }_B}} \right)}}{{{{\overline \gamma  }_B}\left( {{m_B} + {K_{B,i}}} \right)\left( {1 + {{\overline K }_E}} \right)}} \to 0  $,
and $\frac{{{{\overline \gamma  }_E}{m_B}\left( {{m_E} + {K_{E,j}}} \right)\left( {1 + {{\overline K }_B}} \right)}}{{{{\overline \gamma  }_B}{m_E}\left( {{m_B} + {K_{B,i}}} \right)\left( {1 + {{\overline K }_E}} \right)}} \to 0  $.
The Mellin-Barnes integral over $ {\Re _1}$ and $ {\Re _2} $ can be calculated approximately by evaluating the residues at the minimum poles on the right (${r_1} = 1 - {m_E}$, $ {r_2} = {m_E}$) \cite[Theorem 1.11]{kilbas2004h}. With the help of $ \mathop {\lim }\limits_{t \to 0} t\Gamma \left( t \right) = 1 $, we get
 {\small  \begin{align}\label{CNEW2}
&\frac{1}{{2\pi i}}\!\!\!\int_{\!{\Re _{\!1}}}\!\!\!\! {\frac{{\Gamma\! \!\left( \!{2\!\! -\!\! {r_{\!1}}\!\! -\!\! {r_{\!2}}\!\! -\!\! {r_{\!3}}\!\! +\!\! {r_{\!4}}}\! \right)}}{{\Gamma \left( {2 \!-\! {r_1} \!- \!{r_2}} \right)}}\Gamma \!\left(\! {{r_1}} \!\right)}\!
\underbrace {\Gamma\! \left( \!{1 \!\!-\!\! {m_{\!E}}\! \!-\!\! {r_{\!1}}}\! \right)\!\!{{\left(\!\! {\frac{{{{\overline \gamma  }_E}{m_B}\left( {1 + {{\overline K }_B}} \right)}}{{{{\overline \gamma  }_{\!B}}\!\left(\! {{m_{\!B}}\!\! +\!\! {K_{\!B,i}}}\! \right)\!\!\left( \!{1\!\! +\!\! {{\overline K }_{\!E}}} \!\right)}}} \!\!\!\right)}^{\!\!\!\!{r_{\!1}}}}}_{{v_1}\left( {{r_1}} \right)}\!\!\!d{r_{\!1}} \notag\\
& \approx  - {\mathop{\rm Re}\nolimits} {\rm{s}}[{v_1}\left( {{r_1}} \right),1 - {m_E}]\notag\\
& =  - \mathop {\lim }\limits_{{r_1} \to 1 - {m_E}} \left( {{r_1} - 1 + {m_E}} \right){v_1}\left( {{r_1}} \right) \notag\\
&=\! \frac{{\Gamma \!\left( \!{1 \!\!+\!\! {m_{\!E}}\!\! -\!\! {r_{\!2}} \!\!-\!\! {r_{\!3}}\!\! +\!\! {r_{\!4}}}\! \right)}}{{\Gamma \!\left(\! {1\!\! +\!\! {m_{\!E}}\!\! -\!\! {r_{\!2}}}\! \right)}}\Gamma\! \left(\! {1 \!+ \!{m_{\!E}}}\! \right)\!\!{\left(\!\! {\frac{{{{\overline \gamma  }_E}{m_B}\left( {1 + {{\overline K }_B}} \right)}}{{{{\overline \gamma  }_{\!B}}\!\left( {{m_{\!B}}\!\! +\!\! {K_{\!B,i}}} \right)\!\left( {1 \!\!+\!\! {{\overline K }_{\!E}}} \right)}}} \!\!\right)^{\!\!1 \!+\! {m_{\!E}}}}\!\!\!\!\!\!\!\!.
\end{align}}
After substituting the term $ {\Gamma \left( {1 + {m_E} - {r_2} - {r_3} + {r_4}} \right)} $ and $ {\Gamma \left( {1 + {m_E} - {r_2}} \right)} $ in \eqref{CNEW2} into \eqref{CNEW1}, we get
 {\small \begin{align}\label{CNEW3}
&\frac{1}{{2\pi i}}\!\!\int_{\!{\Re _{\rm{2}}}} \!\!\!\!\!{\frac{{\Gamma\! \!\left(\! {1\!\! +\!\! {m_{\!E}} \!\!-\!\! {r_{\!2}}\!\! -\!\! {r_{\!3}}\!\! +\!\! {r_{\!4}}}\! \right)}}{{\Gamma \left( {1\!\! +\!\!{m_E}\!\! -\!\! {r_2}} \right)}}\Gamma\!\! \left( {{r_{\!2}}} \right)}\!\underbrace {\Gamma\! \!\left(\! {{m_{\!E}}\!\! -\!\! {r_{\!2}}} \!\right)\!\!\!{{\left(\!\!\! {\frac{{{{\overline \gamma  }_{\!E}}\!{m_{\!B}}\!\!\left( \!{{m_{\!E}} \!\!+\! \!{K_{\!E,j}}} \!\right)\!\!\left( \!{1 \!\!+\!\! {{\overline K }_{\!B}}} \!\right)}}{{{{\overline \gamma  }_{\!B}}\!{m_{\!E}}\!\!\left( \!{{m_{\!B}}\!\! +\!\! {K_{\!B,i}}}\! \right)\!\!\left( \!{1\!\! +\!\! {{\overline K }_{\!E}}}\! \right)}}} \!\!\!\right)}^{\!\!\!{r_{\!2}}}}}_{{v_{\rm{2}}}\left( {{r_{\rm{2}}}} \right)}\!\!\!\!d{r_{\!2}} \notag\\
& \approx  - {\mathop{\rm Re}\nolimits} {\rm{s}}[{v_{\rm{2}}}\left( {{r_{\rm{2}}}} \right),{m_E}] \notag\\
&=  - \mathop {\lim }\limits_{{r_{\rm{2}}} \to {m_E}} \left( {{r_{\rm{2}}} - {m_E}} \right){v_{\rm{2}}}\left( {{r_{\rm{2}}}} \right)\notag\\
& =\! \Gamma\! \left( {1 \!-\! {r_3}\! +\! {r_4}} \right)\!\Gamma \left( {{m_E}} \right)\!\!{\left(\! {\frac{{{{\overline \gamma  }_E}{m_B}\!\left( {{m_E}\! + \!{K_{E,j}}} \right)\!\!\left( {1 \!+ \!{{\overline K }_B}} \right)}}{{{{\overline \gamma  }_B}{m_E}\!\left( {{m_B} \!+\! {K_{B,i}}} \right)\!\!\left( {1 \!+\! {{\overline K }_E}} \right)}}}\! \right)^{{\!\!\!m_E}}}\!\!\!\!.
\end{align}}
When $ {\overline \gamma  _B} \to \infty   $,
we have $ \frac{{{{\overline \gamma  }_B}\left( {{m_B} + {K_{B,i}}} \right)}}{{{m_B}\left( {1 + {{\overline K }_B}} \right)}} \to \infty.  $
The Mellin-Barnes integral over $ {\Re _4}$ can be calculated approximately by
evaluating the residue at the maximum pole on the left ($ {r_4} = 0 $) \cite[Theorem 1.7]{kilbas2004h}.
After substituting $ {\Gamma \left( {1 - {r_3} + {r_4}} \right)} $  in \eqref{CNEW3} into \eqref{CNEW1}, we get
{\small  \begin{align}\label{CNEW4}
&\frac{1}{{2\pi i}}\!\!\int_{{\!\Re _{\rm{4}}}}\!\!\! {\Gamma \!\left( \!{1\! -\! {r_3}\! +\! {r_4}} \right)}\!\underbrace {\frac{{\Gamma \!\left( {1\! -\! {r_4}} \right)\!\Gamma\! \left( {{r_4}} \right)\!\Gamma \!\left( {{r_4}} \right)}}{{\Gamma \left( {1{\rm{ + }}{r_4}} \right)}}{{\left(\! {\frac{{{{\overline \gamma  }_B}\!\left( {{m_B} \!+\! {K_{B,i}}} \right)}}{{{m_B}\!\left( {1\! +\! {{\overline K }_B}}\! \right)}}} \right)}^{{\!r_4}}}\!\!\!\!\!}_{{v_{\rm{4}}}\left( {{r_{\rm{4}}}} \right)}\!d{r_{\rm{4}}} \notag\\
& \approx {\mathop{\rm Re}\nolimits} {\rm{s}}[{v_{\rm{4}}}\left( {{r_{\rm{4}}}} \right),{\rm{0}}]  \notag\\
& = \mathop {\lim }\limits_{{r_{\rm{4}}} \to {\rm{0}}} \frac{d}{{dr_4^{ - 1}}}\left( {r_4^2{v_4}\left( {{r_4}} \right)} \right)\notag\\
& = \Gamma \left( {1 - {r_3}} \right)\left( {\ln \left( {\frac{{{{\overline \gamma  }_B}\left( {{m_B} + {K_{B,i}}} \right)}}{{{m_B}\left( {1 + {{\overline K }_B}} \right)}}} \right) + \psi \left( {1 - {r_3}} \right)} \right),
\end{align}}
where $\psi \left(  \cdot  \right)  $ is Euler psi function \cite[Eq. (8.360.1)]{gradshteyn2007}.
Substituting \eqref{CNEW2}, \eqref{CNEW3} and \eqref{CNEW4} into \eqref{CNEW1}, we get
 {\small \begin{align}\label{CNEW5}
{H_{\!R}} \!\approx&\! \frac{{\rm{1}}}{{2\pi i}}\!\!\int\limits_{{R_3}}\!\!\! {\Gamma \!\left(\! {{r_{\!3}}}\! \right)\!\Gamma\! \left( {1 \!\!- \!\!{m_{\!B}} \!\!-\!\! {r_{\!3}}} \right)\!\!{{\left( \!\!{\frac{{{m_{\!B}}}}{{{K_{\!B,i}}}}}\!\!\right)}^{ \!\!\!{r_{\!3}}}}}\!\!\!\left(\!\! {\ln\!\! \left(\!\! {\frac{{{{\overline \gamma  }_{\!B}}\!\left( \!{{m_{\!B}}\!\! +\!\! {K_{\!B,i}}}\! \right)}}{{{m_{\!B}}\left( {1\!\! + \!\!{{\overline K }_B}} \right)}}} \!\!\right) \!\!\!+\!\! \psi\! \left(\! {1 \!\!-\!\! {r_{\!3}}}\! \right)} \!\!\right)\!\!\!d{r_{\!3}}.
\end{align}}
Substituting \eqref{CNEW5} into \eqref{24} and with the help of \cite[06.14.07.0005.01]{web}, we derive \eqref{R_appr}.

Following the same  derivation process as $ {R_{i,j,appr}} $, the multivariate Fox's $H$-function $ {H_T} $ in \eqref{25} can be
expressed in terms of the Mellin-Barnes integral  which can be calculated approximately by evaluating the residues.
Then we can easily derive  \eqref{T_appr}.

		\section{Proof of theorem \ref{prop3}}\label{AppendixC}
		\renewcommand{\theequation}{D-\arabic{equation}}
		\setcounter{equation}{0}		
		Substituting \eqref{B-3} and \eqref{B-4} into \eqref{32}, we have
		 {\small \begin{align}
		&{P_{out}}\! =\!\! \int_0^\infty \!\!\! { \left( {{p_E}{f_{RS,E,1}}\left( {{\gamma _E}} \right) + \left( {1 - {p_E}} \right){f_{RS,E,2}}\left( {{\gamma _E}} \right)} \right)}\!\notag\\
		&\times\!\left( {{p_B}\!{F_{RS,\!B,\!1}}\!\!\left( \!{{R_s}\!{\gamma _E} \!+\! {R_s}{\rm{\! - 1}}}\! \right) } \right.\left. {\!\!+\! \left(\! {1\! \!-\!\! {p_B}}\! \right)\!{F_{RS\!,B\!,2}}\!\left(\! {{R_s}\!{\gamma _{\!E}}\!\! +\!\! {R_{\!s}} \!\!-\!\! 1} \right)}\! \right)\!d{\gamma _{\!E}} \notag\\
		& \qquad= {W_{1,1}} + {W_{1,2}} + {W_{2,1}} + {W_{2,2}},
		\end{align}}
			where
	 {\small  \begin{align}\label{C-2}
		{W_{i,j}}\! =\!\! \int_0^\infty \!\!\!\!\! {{Q_{\!B\!,i}}{Q_{\!E\!,j}}\!{f_{R\!S\!,E\!,i}}\!\left({{\gamma _E}} \right)\!\!{F_{R\!S\!,B\!,j}}\!\left( {{R_s}\!{\gamma _{\!E}} \!+\! {R_s}\!{\rm{ - 1}}} \right)\!d{\gamma _E}}.
		\end{align}}
		
		For real values of $ m $, with the help of \cite[Eq. (1.4.8)]{srivastava1985multiple}, the confluent Lauricella hypergeometric function  $ {\Phi _2}\left( . \right) $ can be expressed as	
	 {\small \begin{align}\label{ss1}
		{\Phi _2}\left( {{b_1},{b_2};c;{x_1},{x_2}} \right) = \!\!\!\sum\limits_{{n_1} = {n_2} = 0}^\infty \!\! {\frac{{{{\left( {{b_1}} \right)}_{{n_1}}}{{\left( {{b_{\rm{2}}}} \right)}_{{n_{\rm{2}}}}}}}{{{{\left( c \right)}_{{n_1} + {n_2}}}}}\frac{{x_1^{{n_1}}}}{{{n_1}!}}\frac{{x_2^{{n_2}}}}{{{n_2}!}}}.
		\end{align}}
		Substituting \eqref{9} and \eqref{10} into \eqref{C-2} and with the help of \eqref{ss1} and \eqref{A-8}, we obtain \eqref{ss2} at the bottom of this page.		
				\setcounter{equation}{4}
	Using \cite[Eq. (1.111)]{gradshteyn2007}, we obtain
				{\small  \begin{align}\label{k4}
					{\kappa _{\!4}}\! =\!\!\!\!\!\! &\sum\limits_{{n_4} = 0}^{1\! + {n_{\!1}} \!+ {n_{\!2}}}\!\!\! {{\frac{{R_s^{{n_{\!4}}}{w^{1\! +\! {n_{\!1}}\! +\! {n_{\!2}}\!-\! {n_{\!4}}}}\!\!\left(\! {1\! \!+\!\! {n_{\!1}}\!\! +\!\! {n_{\!2}}}\! \right)!}}{{\left( {{n_4}} \right)!\left( \!{1\!\! + \!\!{n_{\!1}} \!\!+\!\! {n_{\!2}}\!\! -\!\! {n_{\!4}}}\! \right)!}}}}  \!\!\! {\int_{\rm{0}}^\infty \!\!\!\!\!\!\! {\gamma _{\!E}^{{n_{\!4}} \!- \!r}\!\!\exp\!\! \left(\! \!\!{ - \frac{{\left( {1\!\!+\!\! {{\overline K }_{\!E}}} \right)\!{m_{\!E}}}}{{{{\overline \gamma  }_{\!E}}\!\!\left(\! {{m_{\!E}} \!\!+\!\! {K_{\!E,j}}}\! \right)}}\!{\gamma _{\!E}}}\!\!\! \right)\!\!d{\gamma _{\!E}}}} ,
					\end{align}}
       With the aid of \cite[Eq. (3.381.4)]{gradshteyn2007},  the integral in \eqref{k4} can
       be derived as
     {\small  \begin{align}\label{k44}
       {\kappa _{\!4}}\! =\!\!\!\!\! \sum\limits_{{n_4} = 0}^{1\! +\! {n_{\!1}}\! +\! {n_{\!2}}} \!\!\! {{\frac{{R_s^{{n_4}}{w^{1\! + \!{n_{\!1}} \!+\! {n_{\!2}}\! -\! {n_{\!4}}}}\!\left( \!{1\!\! +\!\! {n_{\!1}}\!\! +\!\! {n_{\!2}}}\! \right)!}}{{\left( {{n_{\!4}}} \right)!\left(\! {1 \!\!+\!\! {n_{\!1}}\!\! +\!\! {n_{\!2}} \!\!-\!\! {n_{\!4}}}\! \right)!}}}}  \!\!{ {\left(\!\! {\frac{{\left(\! {1\!\! +\!\! {{\overline K }_{\!E}}} \!\right){m_{\!E}}}}{{{{\overline \gamma  }_{\!E}}\!\left(\! {{m_{\!E}}\!\! +\!\! {K_{\!E,j}}} \!\right)}}} \!\!\right)^{ \!\!\!\!- \!1\! -\! {n_3}\! +\! r}}\!\!\!\!\Gamma\! \left( {\!1 \!\!+\!\!{n_{\!3}}\!\! -\!\! r}\! \right)} .
       \end{align}}
       Substituting \eqref{k44} into \eqref{ss2} and using the definition of Meijer's $ G $-function, we derive the close-form expression of
		$ {W_{i,j}} $ for real values of $ m $ as \eqref{ss3}. For integer values of $m$, substituting \eqref{B-3} and \eqref{B-4} into \eqref{C-2} and with the aid of \cite[Eq. (1.111)]{gradshteyn2007}, we obtain \eqref{C-3} at the bottom of this page.
		\setcounter{equation}{7}
			
		With the aid of \cite[Eq. (3.381.4)]{gradshteyn2007},  $ \kappa _5 $ and $ \kappa _6 $ can be expressed as
		{\small  \begin{align}\label{C-4}
		{\kappa _5} = \bar \rho _{E,j}^{{m_E} - n}\Gamma \left( {{m_E} - n} \right),
		\tag{D-8}
		\end{align}
		\begin{align}\label{C-5}
		{\kappa _6} = {\left( {\frac{{{{\overline \rho  }_{B,i}} + {R_s}{{\overline \rho  }_{E,j}}}}{{{{\overline \rho  }_{E,j}}{{\overline \rho  }_{B,i}}}}} \right)^{ - {m_E} + n - k + t}}\Gamma \left( {{m_E} - n + k - t} \right).
		\tag{D-9}
		\end{align}}	
		Substituting \eqref{C-4} and \eqref{C-5} into \eqref{C-3}, we obtain the closed-form expression of  $ {W_{i,j}} $ for integer value of $m$ as \eqref{344}.

		\section{Proof of theorem \ref{prop4}}\label{AppendixD}
		\renewcommand{\theequation}{E-\arabic{equation}}
		\setcounter{equation}{0}
		Substituting \eqref{7} and \eqref{8} into \eqref{33}, we obtain
	{\small 	\begin{align}
		{P_{nz}} & = \int_0^\infty  {\left( {{p_E}{F_{RS,E,1}}\left( {{\gamma _B}} \right) + \left( {1 - {p_E}} \right){F_{RS,E,2}}\left( {{\gamma _B}} \right)} \right)}\notag\\
		&\quad\times\left( {{p_B}{f_{RS,B,1}}\left( {{\gamma _B}} \right) + \left( {1 - {p_B}} \right){f_{RS,B,2}}\left( {{\gamma _B}} \right)} \right)d{\gamma _B} \notag\\
		& = {D_{1,1}} + {D_{1,2}} + {D_{2,1}} + {D_{2,2}},
		\end{align}}
		where
		{\small  \begin{align}\label{D-2}
{D_{i,j}} = \int_0^\infty  {{Q_{B,i}}{Q_{E,j}}{f_{RS,B,i}}\left( {{\gamma _B}} \right){F_{RS,E,j}}\left( {{\gamma _B}} \right)d{\gamma _B}}.
		\end{align}	}
		For real values of $ m $, substituting \eqref{9} and \eqref{10} into \eqref{D-2}, we can rewrite $ {D_{i,j}} $ as
{\small 		\begin{align}\label{D-3}
&{D_{i,j}}\!\! =\!\! {Q_{B\!,i}}{Q_{E\!,j}}\!\frac{{\left(\! {1\! +\! {{\overline K }_E}} \!\right)\!\!\left( \!{1 \!+\! {{\overline K }_B}} \!\right)}}{{{{\overline \gamma  }_E}{{\overline \gamma  }_B}}}\!{\left(\! {\frac{{{m_E}}}{{{m_E}\! +\! {K_{E\!,j}}}}}\! \right)^{\!{m_{\!E}}}}\!\!{\left( \!{\frac{{{m_B}}}{{{m_B}\! +\! {K_{B\!,i}}}}}\! \right)^{\!{m_{\!B}}}}\notag\\
&\times\!\!\int_0^\infty \!\!\! {{\gamma _B}\exp\! \left(\! { - \frac{{\left( {1\! +\! {{\overline K }_B}} \right)\!{\gamma _B}}}{{{{\overline \gamma  }_B}}}} \!\right)}\! {}_1{F_1}\!\left(\! {{m_B},1;\frac{{{K_{B,i}}\left( {1 + {{\overline K }_B}} \right){\gamma _B}}}{{{{\overline \gamma  }_B}\left( {{m_B} + {K_{B,i}}} \right)}}} \!\right)\!\!\notag\\
&\times\!{\Phi _{2}}\!\!\left(\! \!{1 \!-\! {m_E}\!,\!{m_E};2; \!- \!\frac{{1 \!+\! {{\overline K }_E}}}{{{{\overline \gamma  }_E}}}\!{\gamma _B}, \!-\! \frac{{{m_E}\!\left( {1\! +\! {{\overline K }_E}} \right)}}{{{{\overline \gamma  }_{\!E}}\!\left( {{m_{\!E}} \!+\! {K_{\!E,j}}} \right)}}\!{\gamma _B}} \!\right)\!\!d{\gamma _B}\!.
		\end{align}}
		Using \eqref{A-7} and \eqref{A-8}, we obtain
		{\small \begin{align}\label{D-4}
		&{D_{\!i\!,j}}\!=\!  \frac{{{Q_{B,i}}{Q_{E,j}}\left( {1 \!+\! {{\overline K }_E}} \right)\!\left( {1\! +\! {{\overline K }_B}} \right)}}{{{{\overline \gamma  }_{\!E}}{{\overline \gamma  }_{\!B}}{\Gamma \!\left(\! {1 \!\!-\!\! {m_{\!B}}}\! \right)\!\Gamma\! \left( \!{1\! \!-\!\! {m_{\!E}}}\! \right)\!\Gamma\! \left(\! {{m_{\!E}}} \!\right)}}}{\left(\!\! {\frac{{{m_E}}}{{{m_{\!E}}\! \!+\!\! {K_{\!E,j}}}}} \!\!\right)^{\!\!\!{m_{\!E}}}}\! \!\!{\left(\!\! {\frac{{{m_B}}}{{{m_{\!B}} \!\!+\!\! {K_{\!B,i}}}}}\!\! \right)^{\!\!\!{m_{\!B}}}}\!\!\!\!\!\notag\\
		&\times \!\!\!\frac{1}{{{{\left(\! {2\pi i} \!\right)}^{\!3}}}}\!\!\!\int_{\!{\Re _{\!1}}} \!\!{\int_{\!{\Re _{\!2}}}\!\! {\int_{\!{\Re _{\!3}}}\!\!\!\! {\Gamma\! \left(\! {{r_{\!1}}} \!\right)\!\Gamma\! \left(\! {1 \!\!-\!\! {m_{\!E}} \!\!-\!\! {r_{\!1}}} \!\right)} } }\frac{{\Gamma \!\left(\! {{r_{\!2}}}\! \right)\!\Gamma \!\left(\! {{m_{\!E}}\! \!-\!\! {r_{\!2}}}\! \right)\!\Gamma \!\left(\! {{r_{\!3}}}\! \right)\!\Gamma\!\left(\! {1\! \!-\!\! {m_{\!B}}\! \!-\! \!{r_3}} \right)}}{{\Gamma \left( {2 - {r_1} - {r_2}} \right)\Gamma \left( {1 - {r_3}} \right)}} \notag\\
		&\times{\left(\! {\frac{{1 + {{\overline K }_E}}}{{{{\overline \gamma  }_E}}}} \!\right)^{ \!\!\!- {r_1}}} \!\!{\left(\! {\frac{{{m_E}\left( {1 + {{\overline K }_E}} \right)}}{{{{\overline \gamma  }_E} \left( {{m_E} + {K_{E,j}}} \right)}}}\! \right)^{\!\!\! - {r_2}}}\!\!{\left(\! {\frac{{{K_{B,i}}\left( {1 + {{\overline K }_B}} \right)}}{{{{\overline \gamma  }_B}\left( {{m_B} + {K_{B,i}}}\! \right)}}} \right)^{\!\!\! - {r_3}}}\!\!\!\!
		\notag\\
		&\times\!\!\underbrace {\int_0^\infty  \!\!\!\!\!\!{{\gamma _{\!B}}^{1 \!- \!{r_1} -\! {r_2} -\! {r_3}}\!\exp \!\!\left(\!\! {- \!\frac{{\left( {1 \!+\! {{\overline K }_B}} \right){m_B}}}{{{{\overline \gamma  }_{\!B}}\!\left( {{m_{\!B}} \!+\! {K_{\!B,i}}} \right)}}\!{\gamma _{\!B}}} \!\!\right)\!\!d{\gamma _{\!B}}} }_{{\kappa _7}}\!d{r_3}d{r_2}d{r_1}\!.
		\end{align}}
		With the aid of \cite[Eq. (3.381.4)]{gradshteyn2007}, $ \kappa _7 $  can be written  as
		{\small \begin{align}\label{D-5}
{\kappa _7} \!=\! {\left( {\frac{{\left( {1\! +\! {{\overline K }_B}} \right){m_B}}}{{{{\overline \gamma  }_B}\left( {{m_B} \!+\! {K_{1,B}}} \right)}}} \right)^{{r_1} + {r_2} + {r_3} - 2}}\Gamma \left( {2\! -\! {r_1} \!-\! {r_2}\! - \!{r_3}} \right).
		\end{align}}
		Substituting \eqref{D-5} into \eqref{D-4} and with the help of the definition of the multivariate Fox's $ H $-function \cite[Eq. (28)]{alhennawi2015closed}, we obtain the closed-form expression of  $ {D_{i,j}} $ as \eqref{35}.
		
		For integer values of $m$, substituting \eqref{B-3} and \eqref{B-4} into \eqref{D-2}, we obtain
		{\small \begin{align}\label{D-6}
		{D_{\!i,j}} \!=& {Q_{\!B\!,i}}{Q_{\!E\!,j}}\!\!\!\!\sum\limits_{l = 0}^{{m_{\!E}}\! -\! 1} \!{\sum\limits_{n = 0}^{{m_{\!B}}\! -\! 1}\!\!\! { {\frac{{B_{l,E,j}}{B_{n,B,i}}}{{\overline \rho  _{B\!,i}^{{m_{\!B}}\! -\! n}\Gamma\! \left(\! {{m_{\!B}}\!\! -\!\! n}\! \right)}}} } } \!{\underbrace {\int_0^\infty \!\!\!\!\! {\gamma _B^{{m_{\!B}}\! -\! n\! -\! 1}\!\exp \!\left(\!\! { - \frac{{{\gamma _B}}}{{{{\overline \rho  }_{\!B,i}}}}}\!\! \right)\!\!d{\gamma _{\!B}}} }_{{\kappa _8}} } \notag\\
		& -\! {Q_{B\!,i}}{Q_{E\!,j}}\!\sum\limits_{l = 0}^{{m_{\!E}} \!- \!1} \!{\sum\limits_{n = 0}^{{m_{\!B}} \!- \!1} \!{\sum\limits_{k = 0}^{{m_{\!E}}\! - l \!- 1}\!\!\!\left( {{\frac{{{B_{l,E,j}}{B_{n,B,i}}}}{{\overline \rho  _{B,i}^{{m_B} - n}\overline \rho  _{E,j}^k\Gamma \left( {{m_B} - n} \right)k!}}}} \right. }}\notag\\
		&\times\left. {\!\!\underbrace {\int_0^\infty  \!\!{\gamma _B^{k + {m_B} - n - 1}\!\exp\! \left( { \!-\! \left(\! {\frac{1}{{{{\overline \rho  }_{B\!,i}}}} \!+\! \frac{1}{{{{\overline \rho  }_{E\!,j}}}}} \right)\!{\gamma _B}} \!\right)\!d{\gamma _B}} }_{{\kappa _9}}} \!\right).
		\end{align}}	
Following the same  derivation process as $ {W_{i,j}} $ for integer values of $ m $ in Appendix \ref{AppendixC}, we  can easily obtain the closed-form expression of  $ {D_{i,j}} $ for integer values of $m$ as \eqref{36}.

		\end{appendices}
		
		\bibliographystyle{IEEEtran}
		\bibliography{IEEEabrv,Ref}

		\end{document}